\documentclass[usenatbib]{mnras}

\usepackage{newtxtext,newtxmath}

\usepackage[T1]{fontenc}
\usepackage{ae,aecompl}

\usepackage{graphicx}	
\usepackage{amsmath}	
\usepackage{siunitx}
\usepackage[version=4]{mhchem}
\usepackage[autostyle=true]{csquotes}
\usepackage{subcaption}
\usepackage{xcolor}
\usepackage{footnote}

\title[]{Molecular tracers of planet formation in the atmospheres of hot Jupiters}

\author[R. Hobbs et al.]{
Richard Hobbs,$^{1}$\thanks{E-mail: rh567@cam.ac.uk}
Oliver Shorttle,$^{1,2}$
And Nikku Madhusudhan$^{1}$
\\
$^{1}$Institute of Astronomy, University of Cambridge, Cambridge, CB3 0HA, UK\\
$^{2}$Cambridge Earth Sciences, University of Cambridge, Cambridge CB2 3EQ, UK\\
}

\date{Accepted XXX. Received YYY; in original form ZZZ}

\pubyear{2019}

\begin{document}
\label{firstpage}
\pagerange{\pageref{firstpage}--\pageref{lastpage}}
\maketitle

\begin{abstract}
The atmospheric chemical composition of a hot Jupiter can lead to insights into where in its natal protoplanetary disk it formed and its subsequent migration pathway. We use a 1-D chemical kinetics code to compute a suite of models across a range of elemental abundances to investigate the resultant abundances of key molecules in hot jupiter atmospheres. Our parameter sweep spans metallicities between 0.1x and 10x solar values for the C/H, O/H and N/H ratios, and equilibrium temperatures of 1000K and 2000K. We link this parameter sweep to the formation and migration models from previous works to predict connections between the atmospheric molecular abundances and formation pathways, for the molecules \ce{H2O}, \ce{CO}, \ce{CH4}, \ce{CO2}, \ce{HCN} and \ce{NH3}. We investigate atmospheric \ce{H2O} abundances in eight hot Jupiters  reported in the literature. All eight planets fall within our predicted ranges for various formation models, however six of them are degenerate between multiple models and, hence, require  additional molecular detections for constraining their formation histories. The other two planets, HD 189733~b and HD 209458~b, have water abundances that fall within ranges expected from planets that formed beyond the \ce{CO2} snowline. Finally, we investigate the detections of \ce{H2O}, \ce{CO}, \ce{CH4}, \ce{CO2}, \ce{HCN} and \ce{NH3} in the atmosphere of HD 209458~b and find that, within the framework of our model, the abundances of these molecules best match with a planet that formed between the \ce{CO2} and \ce{CO} snowlines and then underwent disk-free migration to reach its current location.
\end{abstract}


\section{Introduction}
\label{secF:Introduction}

\newcommand{\keyword}[1]{\textbf{#1}}
\newcommand{\tabhead}[1]{\textbf{#1}}
\newcommand{\code}[1]{\texttt{#1}}
\newcommand{\file}[1]{\texttt{\bfseries#1}}
\newcommand{\option}[1]{\texttt{\itshape#1}}


The location where a Jupiter-like planet forms around a star can leave traces upon the chemical composition of its atmosphere. Planets are formed out of the dust and gas surrounding their host star. Therefore, any variations in the composition of the gas or dust in the proto-planetary disk that depend on the distance from the star should still be evident in the planet itself, depending on the ratio of dust to gas it is formed from (e.g.,\citealt{Oberg2011,Helling2014, Madhu2014, Cridland2016, Madhu2016, Mordasini2016, Venturini2016, Booth2017, Eistrup2018, Pudritz2018, Madhu2019, Khorshid2021}). 

At greater radial distances in the disk, various volatile species pass their snowlines, where they transition from the gas phase into being frozen as ice. In particular, oxygen rich species such as \ce{H2O} and \ce{CO2} have their snowlines at higher temperatures (and therefore closer to the star) than \ce{CO}: The water snowline in the disk mid-plane is typically at $\sim$170 K, the \ce{CO2} is at $\sim$ 80 K and the \ce{CO} snowline is at $\sim$30 K (\citealt{Martin2014}). Thus, the solids throughout most of the disk have sub-solar \ce{C/O} ratios, and the gas has a super-solar \ce{C/O} ratio. Furthermore, the C/H and O/H ratios of the gas are sub-solar outside the snowlines of the corresponding molecules. In the same way, the N/H ratio in the gas becomes sub-solar once \ce{NH3} freezes out at $\sim$100 K (\citealt{Martin2014}).

There are two routes through which a gas giant is expected to form; core accretion closer to the star (\citealt{Pollack1996,Lissauer2007}), and gravitational instability further from the star (\citealt{Boss2000,Gammie2001,Boley2009,Boley2010,Forgan2011}). In the gravitational instability scenario the planet forms early and quickly, accreting dust and gas almost simultaneously. As such, such planets are typically expected to have a similar bulk composition to that of the disk as a whole, and therefore similar to the stellar metallicity (\citealt{Boss1997, Helled2010}). However, more recent work has shown that it is possible for planets formed via gravitational instability to have enriched or depleted metalicities based upon the size of the solids being accreted; if most solids are between 0.1m and 100m the resulting planet can become enriched, while if most solids are above 1km the planet may end up depleted (\citealt{Helled2014}). For the core accretion scenario a planet's resulting atmospheric metallicity is a more complicated function of planet location, being dependent on the solid to gas partitioning of elements. Most models of core accretion require several million years to form a rocky core and accrete a gaseous envelope (e.g., \citealt{Pollack1996, Lissauer2007, Kobayashi2011}). The composition of the accreted gas is dependent upon the location of the forming planet within the disk. However, enrichment of the atmosphere by solids can also occur during atmospheric accretion, resulting in a variable overall composition (\citealt{Pollack1996}). 


However, there are challenges to linking gas giant atmospheric composition directly to formation location. The presence of hot Jupiters suggests that giant planets can migrate through their system. This is because gaseous giant planets are not expected to form on such short period orbits (e.g., \citealt{Mayor1995,Wu2003,Papaloizou2007,Wu2011}). The accretion of solids (in the form of planetesimals or core erosion) onto the gas giant can result in a significantly different atmospheric composition, especially if solid accretion occurs during migration. Accretion of solids typically results in planetary compositions that are solar or super-solar in O/H and C/H but sub-solar in C/O ratio, since the solids within the \ce{CO} snow line are  oxygen rich (e.g., \citealt{Oberg2011,Madhu2014,Venturini2016,Mordasini2016}). On the other hand, formation of giant planets beyond the \ce{H2O}, \ce{CO2} or \ce{CO} snowlines without significant solid accretion followed by disk free migration can result in low atmospheric metallicities and high C/O ratios \citep[e.g.][]{Madhu2014}. Additionally, radial drift of icy grains from the outer disk can supply volatiles to the inner disk (\citealt{Booth2017}). This results in the enrichment of volatiles at the snowlines of up to 10$\times$ C/H or O/H. This allows gas giants to become metal rich by directly accreting metal-rich gas, and opens the possibility of planets with super-solar C/H and super-solar C/O. Recently, studies have also investigated how N/H ratios may vary in a planet's atmosphere based upon where the planet forms (\citealt{Bosman2019, Oberg2019,Turrini2021}). This additional parameter may help to break some of the degeneracy when examining the composition of giant planets, refining the estimates of where a particular planet may have formed. 

Some studies have examined how the molecular composition of a hot Jupiter may vary based on its elemental abundances  (\citealt{Madhu2012,Moses2013,Tsai2017}), and others have studied how the elemental abundances may change based on where the hot Jupiter formed and how it migrated \citep[e.g.][]{Madhu2014,Mordasini2016,Cridland2020}. However, very few have tried to link these two regimes such that it becomes possible to predict where a hot Jupiter formed based upon its molecular composition. Those that have attempted to do so typically use local thermochemical equilibrium models (\citealt{Mordasini2016}). However, purely chemical equilibrium models are not adequate to fully represent the atmospheric chemistry of hot Jupiters. 

When considering nitrogen bearing species, thermochemical models of Hot Jupiters with atmospheric temperatures above $700\,\mathrm{K}$ predict \ce{N2} to be the dominant nitrogen bearing molecule at around 1 bar (e.g., \citealt{Moses2011, Venot2012}). \ce{N2} is not detectable by spectral observations, which in principle makes determining the N/H ratio of hot Jupiter atmospheres challenging. However, there have been potential detections of the nitrogen bearing molecules \ce{NH3} and \ce{HCN} in the atmosphere of HD209458b, a planet whose temperature is far above $700\,\mathrm{K}$ (\citealt{Giacobbe2021}). This would only be possible if disequilibrium chemical effects, such as diffusion and photochemistry, were impacting the composition of the planet's atmosphere, as has been expected by some previous works (e.g., \citealt{Macdonald2017,Kawashima2021}). As such, in this work, we compute a full suite of disequilibrium chemical models over the range of C/H, O/H and N/H ratios expected from planet formation models. In doing so we aim to more accurately link the observed molecular abundances of hot Jupiter observations to the planet's elemental composition, and thus to the planet's formation and migration history.  

We acknowledge here some of the limitations that exist within our work. The primary one being that this work is generally meant to present a route that can be taken, with evidence being presented to show its merit, rather than an in-depth model of a single exoplanet. As such, our models use generic physical and atmospheric values, rather than specific values, including choosing to calculate both the P-T profile and $K_{zz}$ profile from first principles instead of using retrieved profiles.

In Section \ref{secF:Section2} we provide details of the chemical kinetics code we are using to create our models, as well as the atmospheric parameters of the hot Jupiters we are modelling. We present the results of these models in Section \ref{secF:Section3}. In Section \ref{secF:Section4} we discuss the results of our models, and compare them with different formation scenarios from previous works. We compare our models to observed molecular abundances in some hot Jupiters in Section \ref{secF:Section5}. We discuss our findings and review our work in Section \ref{secF:Conclusion}.

\section{Methods}

\subsection{The Atmospheric Model} 
\label{secF:Section2}

To calculate the abundance of HCNO species throughout the atmospheres of hot Jupiters, we choose to use a chemical kinetics code that can model the effects of disequilibrium chemistry in these atmospheres. We previously developed a disequilibrium chemical kinetics code, \textsc{Levi}, to model hot Jupiter atmospheres. The full development, testing and benchmarking of this code can be found in \cite{Hobbs2019}. A summary of the salient points from this previous work, and a description of changes made to the model for this work will be provided in this section.

The code, \textsc{Levi}, in this work is being used to model the atmospheric chemistry of Jupiter like planets. It does so by calculating the interactions between chemical species, the effects of vertical mixing due to eddy-diffusion, molecular diffusion and thermal diffusion, and photo-chemical dissociation due to an incoming UV flux. It uses input parameters such as the desired equilibrium temperature of the planet and the planet's radius, profiles for the UV stellar spectrum, the pressure-temperature (P-T) profile of the atmosphere, and the eddy-diffusion ($K_{zz}$) profile. It uses the assumptions of hydro-static equilibrium, the atmosphere being an ideal gas, and that the atmosphere is small compared to the planet, such that gravity is constant throughout the atmospheric range being modelled. 

In this work, as in \cite{Hobbs2019}, we will limit our network to exploring the chemistry of \ce{H}, \ce{He}, \ce{C}, \ce{N}, \ce{O} species. It is possible that species containing other elements could affect the results produced. There are some cases in which sulfur chemistry (\citealt{Zahnle2016,Hobbs2021}) can impact the results of this paper. However, the inclusion of these sulfur species goes beyond the scope of this work.

As is typical for codes of this type, we solve the coupled  one-dimensional continuity equation:

\begin{equation}
\frac{\partial n_{i}}{\partial t} = \mathcal{P}_{i} - \mathcal{L}_{i} - \frac{\partial \Phi_{i}}{\partial z},
\label{eq:continuity}
\end{equation}
where $n_{i}$ (\si{\per\metre\cubed}) is the number density of species $i$, with $i = 1,...,N$, with $N$ being the total number of species. $\mathcal{P}_{i}$ (\si{\per\metre\cubed\per\second}) and $\mathcal{L}_{i}$ (\si{\per\metre\cubed\per\second}) are the production and loss rates of the species $i$. $\partial t$ (\si{\second}) and $\partial z$ (\si{\metre}) are the infinitesimal time step and altitude step respectively. $\Phi_{i}$ (\si{\per\metre\squared\per\second}) is the upward vertical flux of the species, given by,
\begin{equation}
\Phi_{i} = -(K_{zz}+D_i)n_{t}\frac{\partial X_{i}}{\partial z} + D_i n_i\left(\frac{1}{H_0} - \frac{1}{H} - \frac{\alpha_{T,i}}{T}\frac{dT}{dz}\right),
\label{eq:diffusion}
\end{equation}
where $X_{i}$ is the mixing ratio of molecule i, and $n_{t}$ (\si{\per\metre\cubed}) is the total number density of molecules such that $n_{i} = X_{i}n_{t}$. The eddy-diffusion coefficient, $K_{zz}$ (\si{\metre\squared\per\second}), approximates the rate of vertical transport and $D_i$ (\si{\metre\squared\per\second}) is the molecular diffusion coefficient of species i. $H_0$ (\si{\metre}) is the mean scale height of the atmosphere, $H$ (\si{\metre}) is the molecular scale height, $T$ (K) is the temperature, and $\alpha_{T,i}$ is the thermal diffusion factor. For the full explanation of how we determine each of these parameters, and solved the equations, see \cite{Hobbs2019}.

Unlike in the previous work, we no longer use precalculated $K_{zz}$ profiles, instead we calculate the profile iteratively and self-consistently using the equations described in \cite{Zhang2017} and \cite{Komacek2019}:

\begin{equation}
    K_{zz} \sim \frac{W^2}{\tau_{\textrm{chem}}^{-1}+\frac{W}{H}},
\end{equation}
where $H=RT/g$ is the scale height of the atmosphere, with $R$ being the gas constant, $T$ being the temperature at the pressure being calculated and $g$ being the gravitational acceleration of the planet. $\tau_{\textrm{chem}}$ is the timescale of chemical interactions, and is equal to:
\begin{equation}
    \tau_{\textrm{chem}} = \frac{[X]}{d[X]/dt},
\end{equation}
where $[X]$ is the abundance of some molecule X. For our calculation of the $K_{zz}$ profile, we use an approximation for the chemical timescale such that it is equal to the average timescale of chemical interactions at each pressure in the atmosphere. This simplifies the $K_{zz}$ profile to a single curve rather than an independent profile for each species in the atmosphere. $W$ is the vertical wind speed, where $\frac{W}{H} \sim \frac{U}{a}$. Here, $a$ is the radius of the planet, and $U$ is the characteristic horizontal wind-speed, defined as 
\begin{equation}
    U \sim \frac{2 \gamma U_{\textrm{eq}}}{\alpha + \sqrt{\alpha^2 + 4 \gamma^2}},
\end{equation}
where $U_{eq}$ is the speed of the maximum cyclostophic wind, defined as
\begin{equation}
    U_{\textrm{eq}} = \frac{a}{\tau_{\textrm{adv,eq}}},   
\end{equation}
and the dimensionless parameters $\alpha$ and $\gamma$ are defined as
\begin{equation}
    \alpha = 1 + \frac{\Omega\, \tau_{\textrm{wave}}^2}{\tau_{\textrm{rad}}\, \Delta \textrm{ln}\, p},   
\end{equation}
\begin{equation}
    \gamma = \frac{\tau_{\textrm{wave}}^2}{\tau_{\textrm{rad}}\,\tau_{\textrm{adv,eq}}\, \Delta \textrm{ln}\, p}.   
\end{equation}

$\Omega$ is the planet's rotation rate, set equal to the orbital period in this work since we assume hot Jupiters are always tidally locked. $\Delta \textrm{ln}\, p$ is the pressure difference between the pressure of interest and the deep atmosphere, where the day and night temperatures are the same. We set our deep atmosphere to be below 10 bars (\citealt{Komacek2017}). In the deep atmosphere, we set $K_{zz} = 10^6 \mathrm{m}^2 \mathrm{s}^{-1}$ using an approximation for the adiabatic region from \cite{Stone1976}. The timescales in the above equations are: The Kelvin wave propagation time across a hemisphere;
\begin{equation}
   \tau_{\textrm{wave}} = a/NH,  
\end{equation}
where $N$ is the Brunt-V\"{a}is\"{a}l\"{a} frequency, the radiative timescale;
\begin{equation}
   \tau_{\textrm{rad}} = \frac{p\, c_p}{4\,g\,\sigma\,T^3},  
\end{equation}
where $c_p$ is the specific heat capacity of the atmosphere, and $\sigma$ is the Stefan-Boltzmann constant. Lastly, the advective timescale that a cyclostrophic wind induced by the day-night temperature difference in radiative equilibrium would have is;
\begin{equation}
   \tau_{\textrm{adv,eq}}= a \sqrt{\frac{2}{R\, k_b\, T_{eq} \, \Delta \textrm{ln}\, p}},  
\end{equation}
where $k_b$ is the Boltzmann constant and $T_{eq}$ is the equilibrium temperature of the planet.

We also use a variable parametisation for the P-T profile of the atmosphere using the equations described in \cite{Guillot2010}. The temperature of the atmosphere is parametised as;

\begin{multline}
    T^4 = \frac{3}{4}\,T_{\textrm{int}}^4\,(\frac{2}{3} + \tau) +\\ \frac{3}{4}\,T_{\textrm{irr}}^4\,f\,\left(\frac{2}{3} + \frac{\mu_*}{\gamma} + (\frac{\gamma}{3\,\mu_*} - \frac{\mu_*}{\gamma}) \, \textrm{exp}(-\frac{\gamma\, \tau}{\mu_*})\right).
\end{multline}

Here, $\mu_* = \textrm{cos}\, \theta_* = 1/\sqrt{3}$, where $\theta_*$ is the incidence angle for stellar radiation, $\gamma = \kappa_{\textrm{v}}/\kappa_{\textrm{th}}$ is the ratio between the mean visual ($\kappa_{\textrm{v}}$) and thermal ($\kappa_{\textrm{th}}$) opacities, and $\tau$ is the optical depth. We chose the values for the visual and thermal opacities to be $\mathrm{4x10^{-3}\, cm^2\, g^{-1}}$ and $\mathrm{10^{-2}\, cm^2\, g^{-1}}$ respectively, approximately the values for HD 209458b (\citealt{Guillot2010}). $f=1/2$ is a flux factor for isotropic radiation averaged over the day-side of the planet. The interior temperature is set to be $T_{\textrm{int}}=300\,\textrm{K}$ and the irradiation temperature is $T_{\textrm{irr}} = \sqrt{2}\, T_{\textrm{eq}}$, where $T_{\textrm{eq}}$ is the equilibrium temperature of the planet. For $T_{\textrm{eq}}$ we assume the planet has an albedo of 0 and efficient energy redistribution. 

 We use the values for solar metallicity from \cite{Asplund2009}. This gives an elemental ratio, as a fraction of the total number of molecules, of: $X_{\ce{H2}}=0.5\times X_{\ce{H}}=0.8535$, $X_{\ce{He}}=0.145$, $X_{\ce{C}}=4.584\times10^{-4}$, $X_{\ce{O}}=8.359\times10^{-4}$, $X_{\ce{N}}=1.154\times10^{-4}$. In our models, we independently alter the metallicity between 0.1x and 10x these solar values for each of \ce{C}, \ce{O}, \ce{N}. \ce{He} is kept constant, and \ce{H2} is altered such that $X_{\ce{H2}}$, $X_{\ce{He}}$, $X_{\ce{C}}$, $X_{\ce{O}}$ and $X_{\ce{N}}$ sum to unity.
 
 The equilibrium temperatures of the modelled hot Jupiters range across five temperatures between $1000\,\mathrm{K}$ and $2000\,\mathrm{K}$, although we only show the two extremes in this work. To produce a suite of models over the full range of bulk compositions, we ran our model for each point in a 9x9x9 grid in the C/H, O/H and N/H parameter space, for each planetary equilibrium temperature being investigated. This produced a total of 729 models per modelled planet, with data points for X/H equally spaced in logspace between 0.1x and 10x the solar values for X/H. 
 
 It is worth noting that due to our prescription for $K_{zz}$, the different equilibrium temperatures of our hot Jupiter models will result in different $K_{zz}$ values. While this masks the exclusive effect of temperature as a variable in setting the chemistry of hot Jupiter atmospheres, our primary aim is to model realistic variations in hot Jupiter atmospheric chemistry due to compositional differences. Therefore, a sweep of orbital radius (and therefore temperature and $K_{zz}$), explores the range of hot Jupiter atmosphere chemistry for a given composition.
 
Except for orbital radius (and thus equilibrium temperature) and metallicity, we keep all other planetary and stellar parameters constant throughout all of our models, except where explicitly stated otherwise. These constant parameters are listed in Table \ref{tab:param}. The UV spectrum we apply to our models is the solar spectrum, scaled based upon the distance between the star and the planet.

\begin{table}
    \centering
    \begin{tabular}{c|c}
    \hline

        Planetary gravity   &  10   $\mathrm{m s^{-2}}$ \\ 
        Planetary radius        & 1    $\mathrm{R_J}$  \\
        Planet orbital radius (1000K)      & 0.31    AU  \\
        Planet orbital radius (2000K)       & 0.077   AU  \\
        Stellar radius       & 1     $\mathrm{R_\odot}$ \\
        Stellar temperature       & 5780    $\mathrm{K}$  \\
        Stellar mass       & 1    $\mathrm{M_\odot}$  \\
        \hline

    \end{tabular}
    \caption[Plantary and Stellar parameters]{The planetary and stellar parameters kept constant throughout all of our models.}
    \label{tab:param}
\end{table}
 
 \subsection{Planet composition models}
 \label{secF:Section2.1}
 
 In this section we discuss how different formation and migration mechanisms may lead to different metalicities in the atmosphere of a hot Jupiter. Throughout this work we use metalicity to refer to the full suite of C/H, O/H and N/H ratios, and we use the convention that X/H refers to the absolute value, while [X/H] refers to a value normalised to the solar abundances. We use the values for solar from \cite{Asplund2009}. The following formation and migration pathways came from three different works; \cite{Madhu2014}, \cite{Booth2017} and \cite{Turrini2021}.
 
 The work of \cite{Madhu2014} does not consider nitrogen composition, however, they do consider the widest range of formation scenarios of the studies we draw from. They model hot Jupiters that formed via core accretion and then migrated via both in-disk and disk-free migration, as well as hot Jupiters that formed via gravitational instability before migrating disk-free. This produces a wide range of potential C/O ratios for a hot Jupiter, based upon its history.

\cite{Madhu2014} found that hot Jupiters formed by core accretion between 2 AU and 20 AU before undergoing disk migration had 1 < [O/H] < 10, and 1 < [C/H] < 5, with a C/O ratio that was always sub-solar. Hot Jupiters that formed by core accretion but then underwent disk-free migration fall into two regimes. Those that formed closer in have slightly super-solar C/H and O/H, up to [C/H] =2 and [O/H]=4, but still sub-solar C/O, while those that formed further away, beyond the \ce{CO2} snowline, had sub-solar C/H and O/H, down to [C/H]=0.6 and [O/H]=0.4, but a super-solar C/O ratio. Lastly, planets that formed beyond the \ce{CO2} snowline but within the \ce{CO} snowline by gravitational instability and then migrated inwards also tend to have either super-solar metalicity but a sub-solar C/O ratio or sub-solar metalicity with a super-solar C/O ratio. However, if they formed beyond the \ce{CO} snowline, beyond $\sim$100 AU, they can be any metalicity within our parameter space, but at a solar C/O ratio.

The work of \cite{Booth2017} also does not consider nitrogen in their models. However, their models of chemical enrichment by pebble drift result in compositions in regions of the parameter space that were forbidden by previous works. Through the accretion of metal rich gas the composition of a gas giant could end up with a [C/H] ratio up to 5, and a C/O ratio between solar (0.55) and 1. For Jupiter mass planets forming within the \ce{CO2} snowline, \cite{Booth2017} tend to find metalicities of [O/H] = 2 and between 2 < [C/H] < 3, thus producing super-solar C/O ratios in these high metalicity planets. For planets forming beyond the \ce{CO2} snowline, most formation locations result in a C/O ratio of 1, along the entire parameter space of metalicities.

\cite{Turrini2021} do not consider as wide a range of migration routes as the previous works we have compared to, but they do include the nitrogen in their models as a possible way of breaking the degeneracy arising from consideration of only the C/O ratio. They consider 6 different formation locations for a hot Jupiter that forms via core accretion and subsequently migrates through the disk to an orbital radius of 0.04 AU, accreting solids along the way. How these formation locations relate to the atmospheric C/H, O/H and N/H ratios compared to their solar values is summarised in Table \ref{tab:Turrinibulk}. While we acknowledge that our planetary models orbit at slightly wider radii (between 0.08 AU and 0.3 AU) compared to the 0.04 AU in the work of \cite{Turrini2021}, we assume that any change in the atmospheric composition due to these small differences in migration distance will be sufficiently small to ignore.

\begin{table}
    \centering
    \begin{tabular}{c|c|c|c}
    \hline

        a (AU)   & [C/H]  & [O/H]  &  [N/H]\\ \hline
        5        & 0.93   & 1.06   &  1.02 \\
        12       & 1.33   & 1.28   &  1.09 \\
        19       & 1.77   & 1.85   &  1.35 \\
        50       & 3.23   & 3.70   &  1.89 \\
        100      & 4.67   & 5.19   &  2.43 \\
        130      & 7.33   & 8.33   &  3.65 \\
        \hline

    \end{tabular}
    \caption[Expected bulk composition for different locations of where a hot Jupiter formed]{A summary of the expected bulk compositions of a hot Jupiter from \protect\cite{Turrini2021}.  The planet starts migrating from 6 different formation locations, accreting solids as it travels inwards until its final position at 0.04AU. }
    \label{tab:Turrinibulk}
\end{table}

\begin{figure*}
        \centering
        \includegraphics[width=1\textwidth]{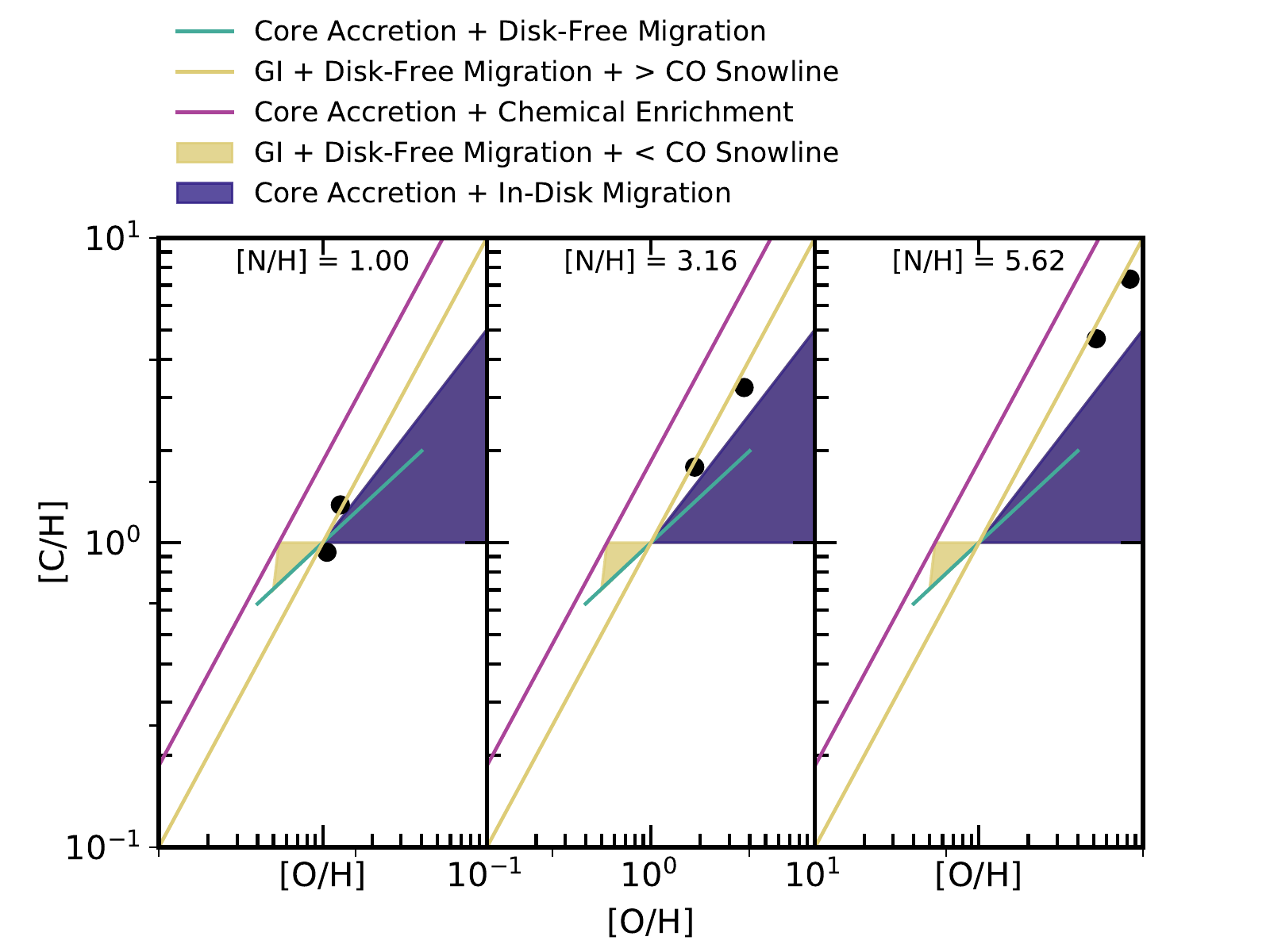}
    \caption[Metalicities expected from different formation models]{The different ranges of metalicities expected from several formation models. The indigo, cyan and yellow lines cover the core accretion and gravitational instability models from \protect\cite{Madhu2014}, the purple lines covers the pebble accretion and enrichment models from \protect\cite{Booth2017}, and the black dots are the metalicities expected from the six formation locations for core accretion in \protect\cite{Turrini2021}. Gravitational instability from within the CO snowline also covers the entire metalicity range of Core Accretion with in-disk migration, but can't be easily seen due to overlap. }
     \label{fig:Form_Metal}
\end{figure*}

A summary of the different metalicities expected from these formation models can be seen in Figure \ref{fig:Form_Metal}.

\section{Model Results}
\label{secF:Section3}

\begin{figure}
        \centering
        \includegraphics[width=0.475\textwidth]{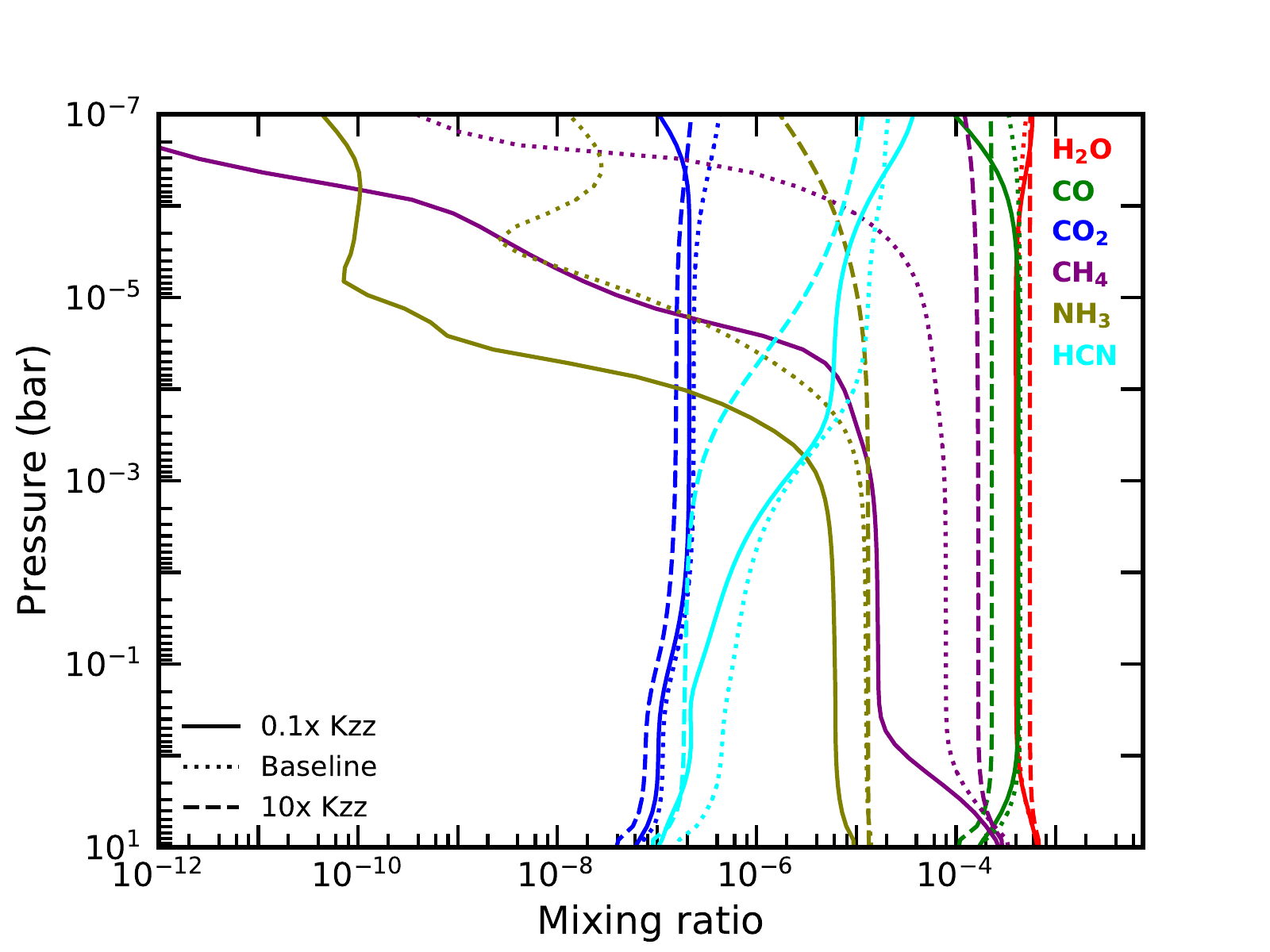}
    \caption[]{A comparison of the abundances of six molecules in the atmosphere of a solar composition hot Jupiter with an equilibrium temperature of $1000\,\mathrm{K}$ across three values of vertical mixing strength.}
     \label{fig:Kzzcheck}
\end{figure}

\begin{figure}
        \centering
        \includegraphics[width=0.475\textwidth]{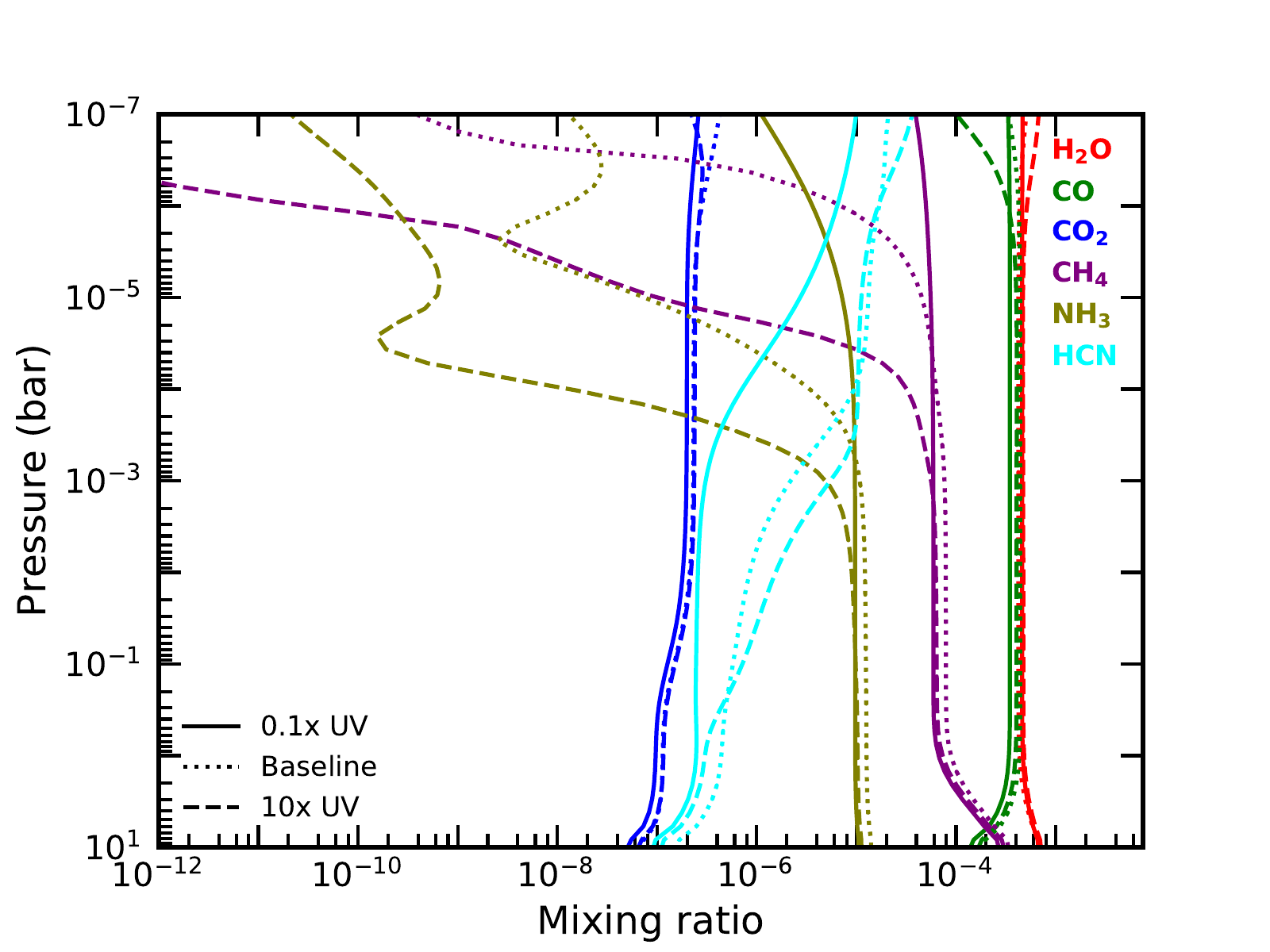}
    \caption[]{A comparison of the abundances of six molecules in the atmosphere of a solar composition hot Jupiter with an equilibrium temperature of $1000\,\mathrm{K}$ across three values of UV irradiation strength.}
     \label{fig:UVcheck}
\end{figure}

\begin{figure}
        \centering
        \includegraphics[width=0.475\textwidth]{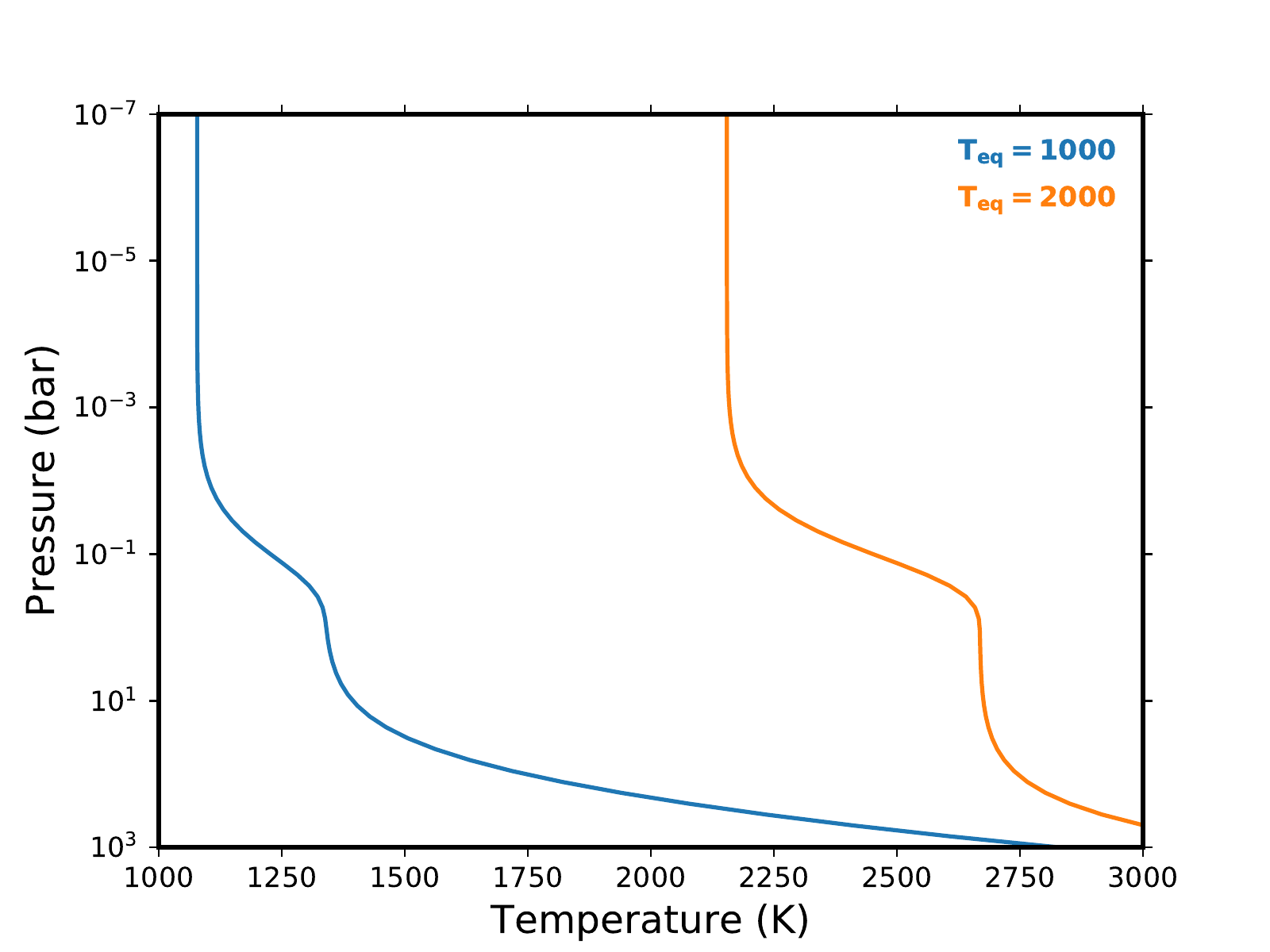}
    \caption[P-T Profiles for hot Jupiters with $\mathrm{T_{eq}}$ = $1000\,\mathrm{K}$ and $2000\,\mathrm{K}$]{The P-T profiles being used in this work. Both were created using the expression presented in Section \ref{secF:Section2}, with an equilibrium temperature of $1000\,\mathrm{K}$ or $2000\,\mathrm{K}$.}
     \label{fig:PT}
\end{figure}

\begin{figure}
        \centering
        \includegraphics[width=0.475\textwidth]{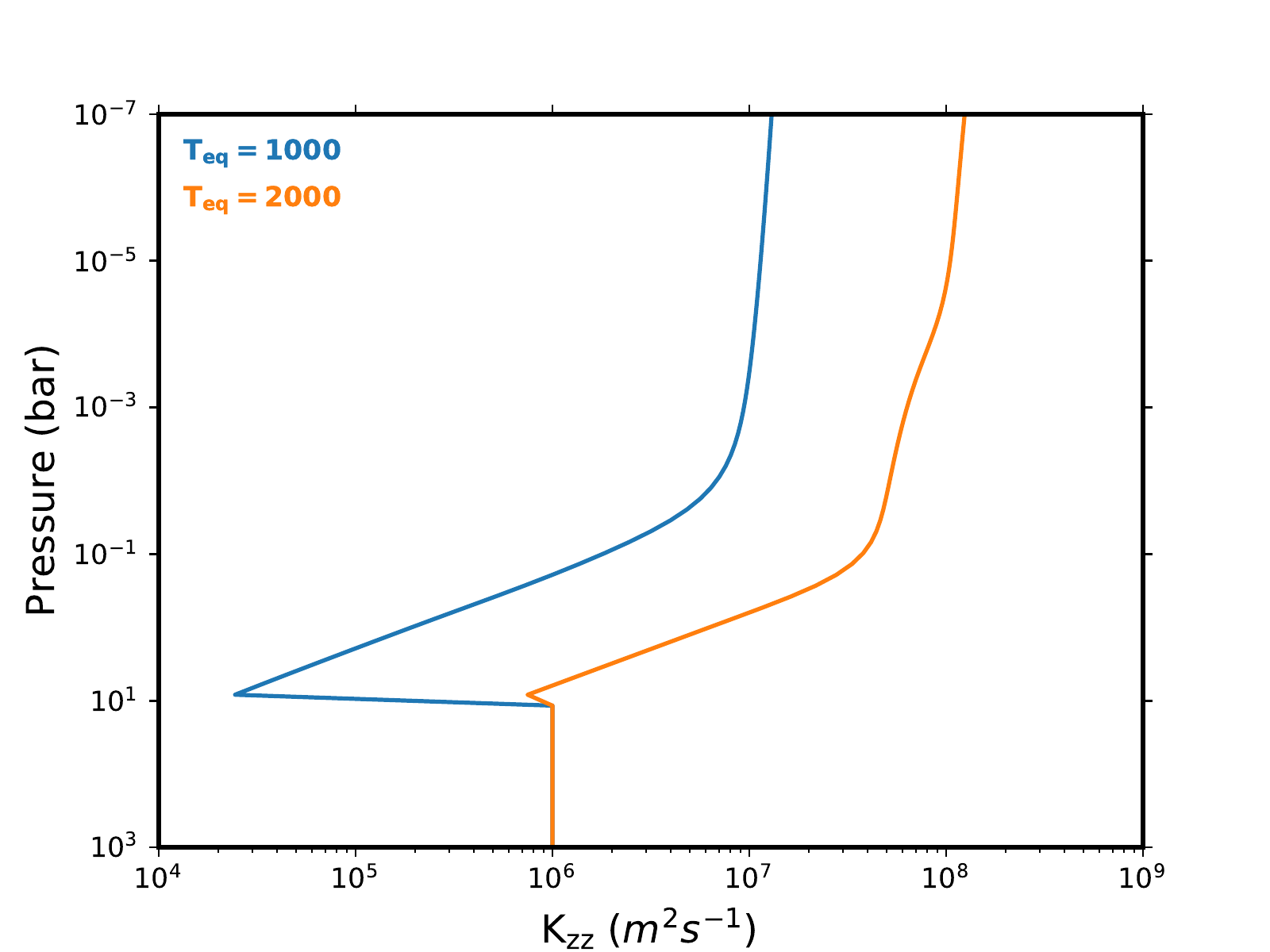}
    \caption[Kzz Profiles for hot Jupiters with $\mathrm{T_{eq}}$ = $1000\,\mathrm{K}$ and $2000\,\mathrm{K}$]{The $K_{zz}$ profiles being used in this work. Both were created using the expression presented in Section \ref{secF:Section2}, with an equilibrium temperature of $1000\,\mathrm{K}$ or $2000\,\mathrm{K}$.}
     \label{fig:Kzz}
\end{figure}

In this section we present the results of our suite of chemical models. We compute models over a grid of C/H, O/H and N/H ratios for two hot Jupiters. We discuss example hot Jupiters at two limiting equilibrium temperatures; $1000\,\mathrm{K}$ and $2000\,\mathrm{K}$. These planets correspond to an orbital radius of 0.3 AU and 0.08 AU around a Sun-like star respectively, assuming an albedo of 0 and efficient energy redistribution. We show the resultant P-T profiles for these planets in Figure \ref{fig:PT} and the $K_{zz}$ profiles in Figure \ref{fig:Kzz}. Additionally, in Figures \ref{fig:Kzzcheck} and \ref{fig:UVcheck} we present variations of the strength of the vertical mixing and incident UV by an order of magnitude in the atmosphere of a solar composition hot Jupiter with a $1000\,\mathrm{K}$ equilibrium temperature, to justify our choice of these values in the rest of this work. In Figure \ref{fig:Kzzcheck} we see that \ce{CH4}, \ce{HCN} and \ce{NH3} are all sensitive to variations in the strength of the vertical mixing, with all three abundances varying across around an order of magnitude at $10^{-3}$ bar. However, in every case, the difference in abundance between the baseline case we use in this work and the case where $K_{zz}$ is 10x stronger is very small. Thus, while we might overestimate the abundance of these three molecules, if the vertical mixing is weaker than we modelled, it is unlikely that we would have underestimated these molecules' abundances. As can be seen in Figure \ref{fig:UVcheck}, only \ce{HCN} is strongly sensitive to UV irradiation, varying by nearly two orders of magnitude at $10^{-3}$ bar. Thus, we must take into account that around strongly irradiating stars that the abundance of \ce{HCN} could be significantly higher than we would predict.

Results of the parameter sweep are shown in Figures \ref{fig:H2O} - \ref{fig:H3N}, presented below. In these figures, we show the C/O ratios of 0.25, 0.54 (solar) and 1 as lines on the figures to assist in determining the abundances at these values. Additionally, we split each figure into three regimes; abundances above $10^{-6}$ which should be detectable, abundances between $10^{-6}$ and $10^{-10}$ which we may one day be able to detect and abundances below $10^{-10}$ which we never expect to be detectable (\cite{Greene2016}). All these results are shown for a pressure of $10^{-3}$ bar, the pressure at which observations of the molecules we model are sensitive to in hot Jupiter atmospheres (\cite{Madhu2019}). Some observations may come from deeper into the planets atmosphere, but as Figures \ref{fig:Kzzcheck} and \ref{fig:UVcheck} show, most molecules have a mixing ratio that is almost unchanging with pressure below $10^{-3}$ bar, and so would not show significant differences in their measured abundances.

\subsection{\texorpdfstring{\ce{H2O}}{H2O} abundance}
\begin{figure}
        \centering
        \includegraphics[width=0.475\textwidth]{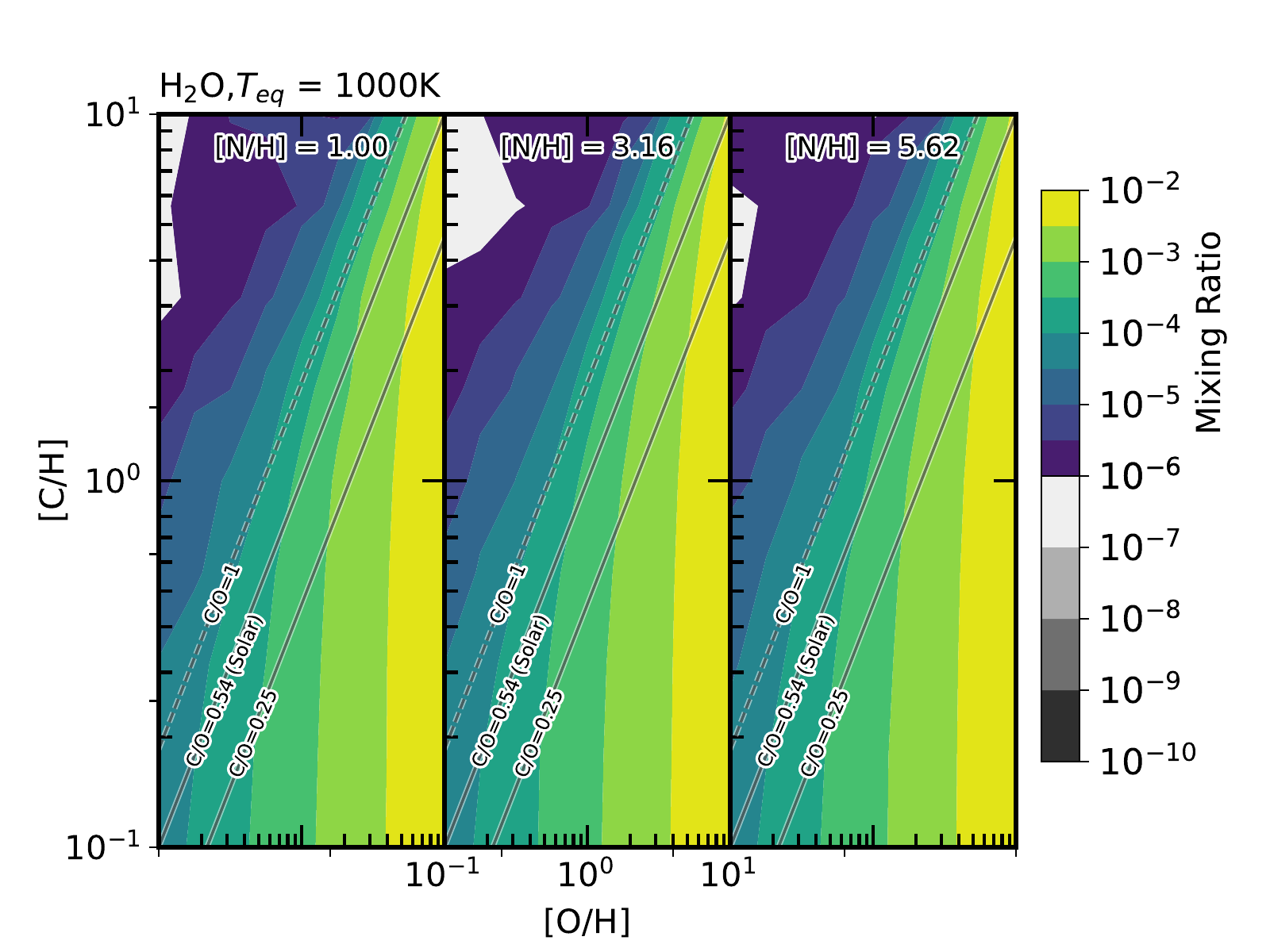}
        \includegraphics[width=0.475\textwidth]{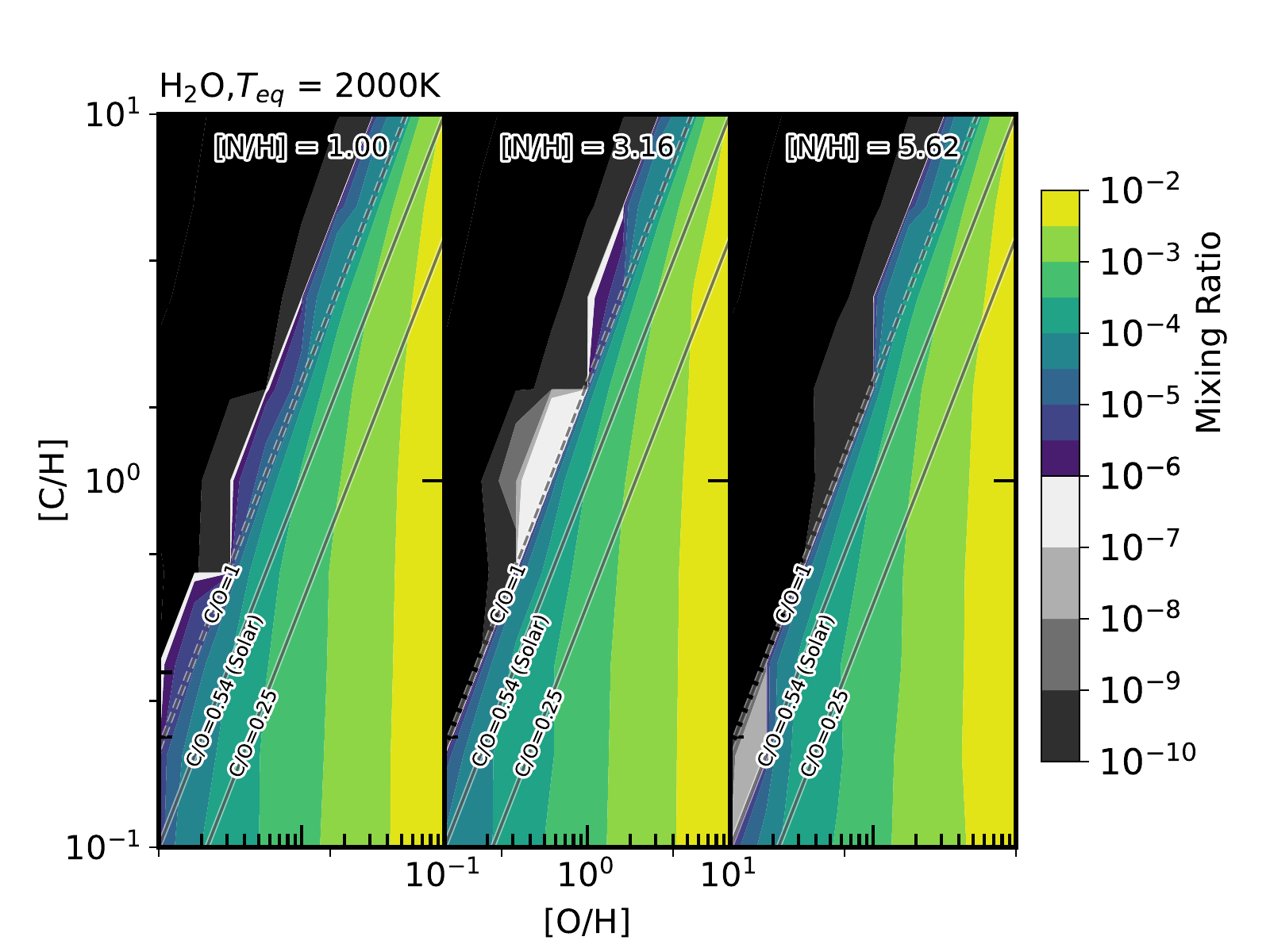}
    \caption[Water abundance across the parameter space]{The abundance of \ce{H2O} for two hot Jupiters, one with a $1000\,\mathrm{K}$ equilibrium temperature (top) and one with a $2000\,\mathrm{K}$ equilibrium temperature (bottom). The variation in abundance of \ce{H2O} is shown against the atmospheric C/H and O/H ratios normalised to solar values. The C/O ratios of 0.25, 0.54 (solar) and 1 are shown on each plot to assist visualisation. Additionally, we consider three atmospheric N/H ratios normalised to solar; 1, 3.2 and 5.6. These are shown from left to right on the figures above.}
     \label{fig:H2O}
\end{figure}

In Figure \ref{fig:H2O} we show how the abundance of water varies over our chosen parameter space. Water's abundance is primarily determined by O/H, however for C/O ratio values greater than 1, a decrease in the abundance of water can be seen (\citealt{Madhu2012,Moses2013}). This is because water does have a weak dependence on the C/H ratio; as the C/O ratio approaches 1, the fraction of O in \ce{CO} becomes increasingly significant, leaving less O to form \ce{H2O}. While C/O < 1, we see a range in \ce{H2O} abundances between $10^{-5}$ at [O/H] = 0.1 and $10^{-2}$ at [O/H] = 10 for both temperature models. For C/O ratios greater than 1, the \ce{H2O} abundance on the $1000\,\mathrm{K}$ hot Jupiter slowly decreases down to $10^{-6}$ as the C/O ratio increases, while on the $2000\,\mathrm{K}$ hot Jupiter, the \ce{H2O} drops by several orders of magnitude immediately.

For planets with a C/O ratio close to 1, \ce{H2O} is a good measure of the C/O ratio in our two hot Jupiter models. This is because the \ce{H2O} becomes increasingly dependent on the C/O ratio once the ratio approaches or exceeds 1. For C/O ratios that exceed 1 in our 2000K model, the abundance of \ce{H2O} would drop below detectable limits, and would not contribute significantly to the measured C/O ratio. However, water can still act as a diagnostic for the C/O ratio, even for C/O ratios great than 1, since the lack of water is itself a diagnostic. At C/O ratios less than 0.25, the \ce{H2O} abundance tends to depend only on the O/H ratio, becoming a worse measure of the C/O ratio. As expected for \ce{H2O}, it is independent of the N/H ratio. There are no N-based species that contain O that are of sufficient abundance to impact the sequestration of O in \ce{H2O}. Thus, \ce{H2O} has no use in determining the N/H ratio.

\subsection{\texorpdfstring{\ce{CO}}{CO} abundance}
\begin{figure}
        \centering
        \includegraphics[width=0.475\textwidth]{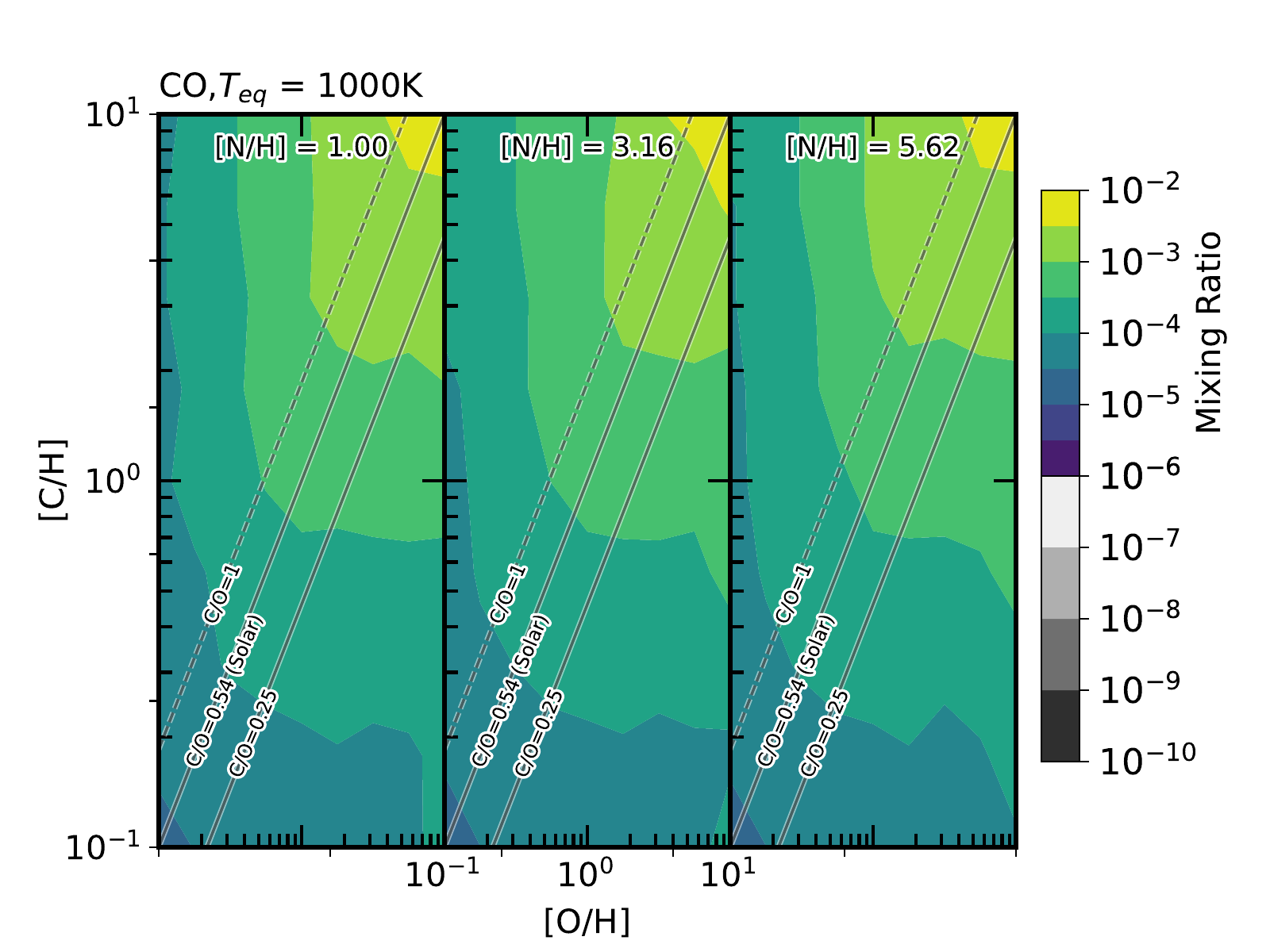}
        \includegraphics[width=0.475\textwidth]{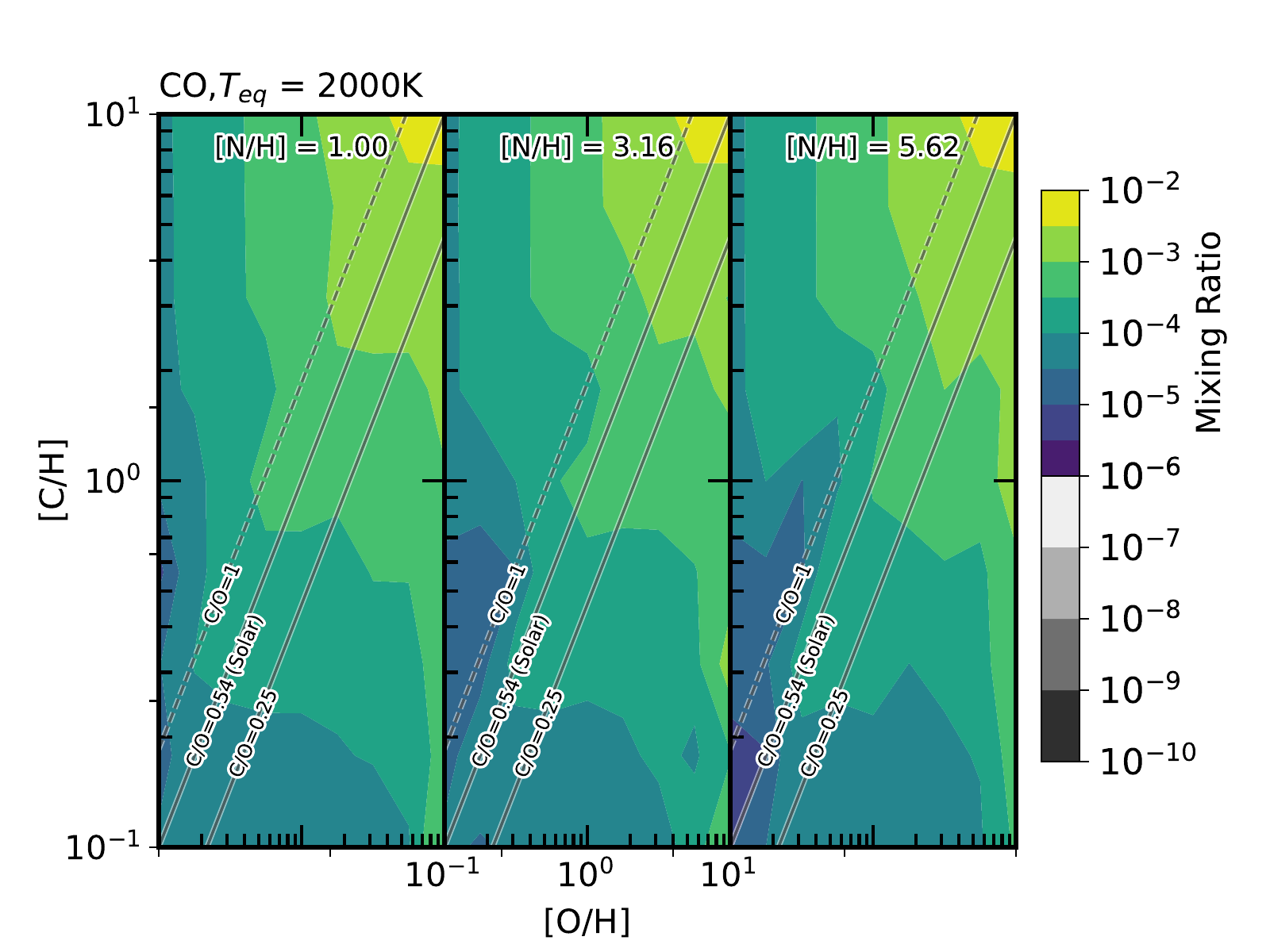}
    \caption[\ce{CO} abundance across the parameter space]{The abundance of \ce{CO} for two hot Jupiters, one with a $1000\,\mathrm{K}$ equilibrium temperature (top) and one with a $2000\,\mathrm{K}$ equilibrium temperature (bottom). The variation in abundance of \ce{CO} is shown against the atmospheric C/H and O/H ratios normalised to solar values. The C/O ratios of 0.25, 0.54 (solar) and 1 are shown on each plot to assist visualisation. Additionally, we consider three atmospheric N/H ratios normalised to solar; 1, 3.2 and 5.6. These are shown from left to right on the figures above.}
     \label{fig:CO}
\end{figure}

Figure \ref{fig:CO} shows how the abundance of \ce{CO} varies across the composition parameter space. \ce{CO} is directly dependent upon both the O/H and C/H ratios. However, the way it is dependent can be quite useful for determining planetary composition. For a C/O ratio less than 1, the \ce{CO} abundance is near independent of the O/H ratio, while for a C/O ratio greater than 1, the \ce{CO} abundance is near independent of the C/H ratio. This is because for C/O < 1, \ce{CO} is the primary carbon carrier, and as long as there is more oxygen than carbon, increasing the amount of oxygen does not assist in creating more \ce{CO}. This relationship is reversed for ratios of C/O > 1. Since there are very few formation models that result in a C/O > 1, we can see that \ce{CO} provides a good measure of the C/H ratio, especially considering that it is predicted to be one of the most abundant species in hot Jupiter atmospheres. In our modelled atmospheres, we find that the abundance of \ce{CO} varies between $10^{-5}$ for [O/H] = 0.1 and [C/H] = 0.1, and $10^{-2}$ for [O/H] = 10 and [C/H] = 10. Once again, we see little effect of the N/H ratio on the abundance of \ce{CO}. While molecules like \ce{HCN} could theoretically take more of the available carbon as the N/H ratio increases, we find that even with [N/H] = 10, the \ce{HCN} abundance is still a small fraction of the \ce{CO} abundance and thus doesn't diminish the atmospheric \ce{CO} reservoir in any significant way.

\subsection{\texorpdfstring{\ce{CH4}}{CH4} abundance}
\begin{figure}
        \centering
        \includegraphics[width=0.475\textwidth]{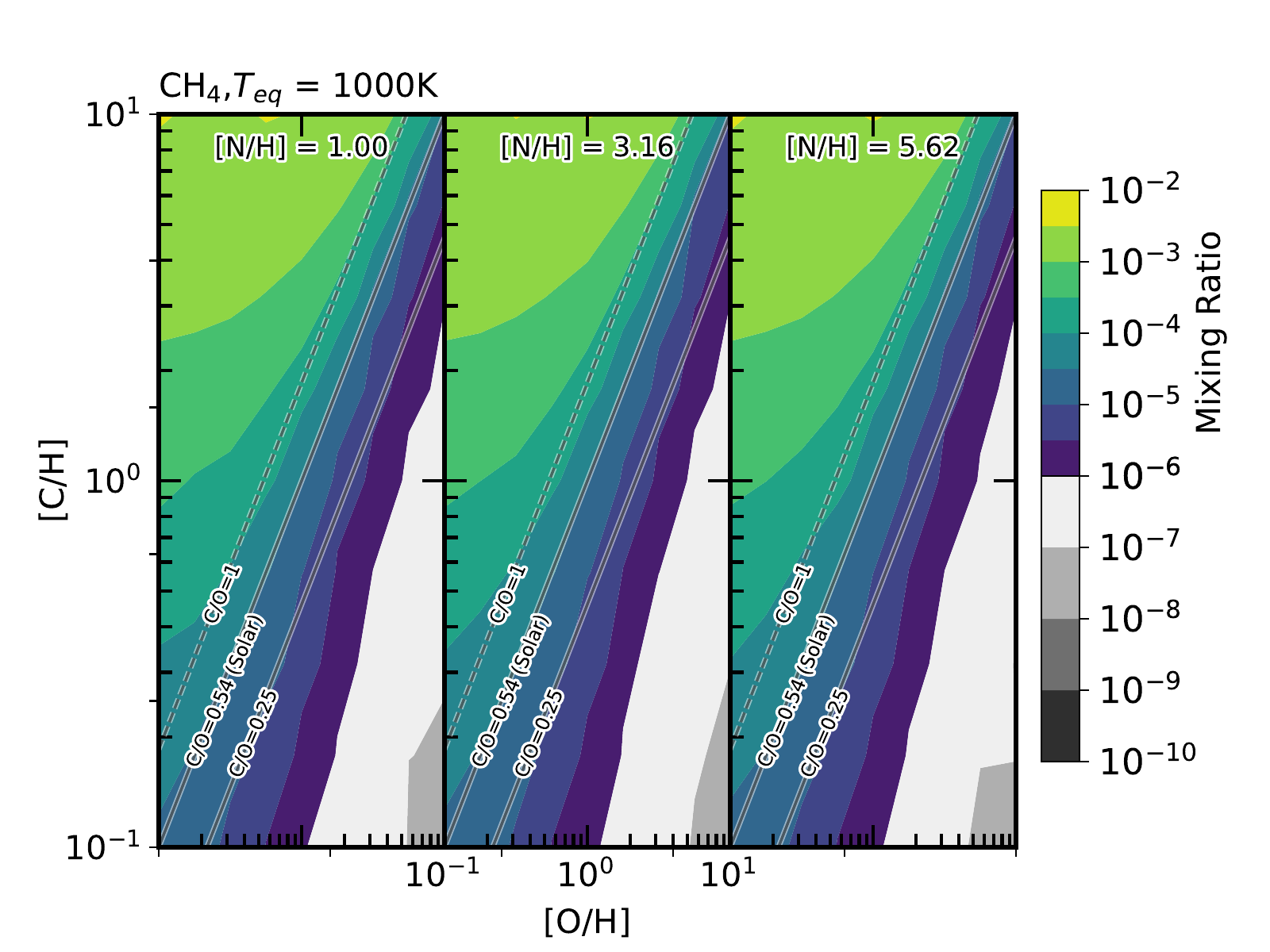}
        \includegraphics[width=0.475\textwidth]{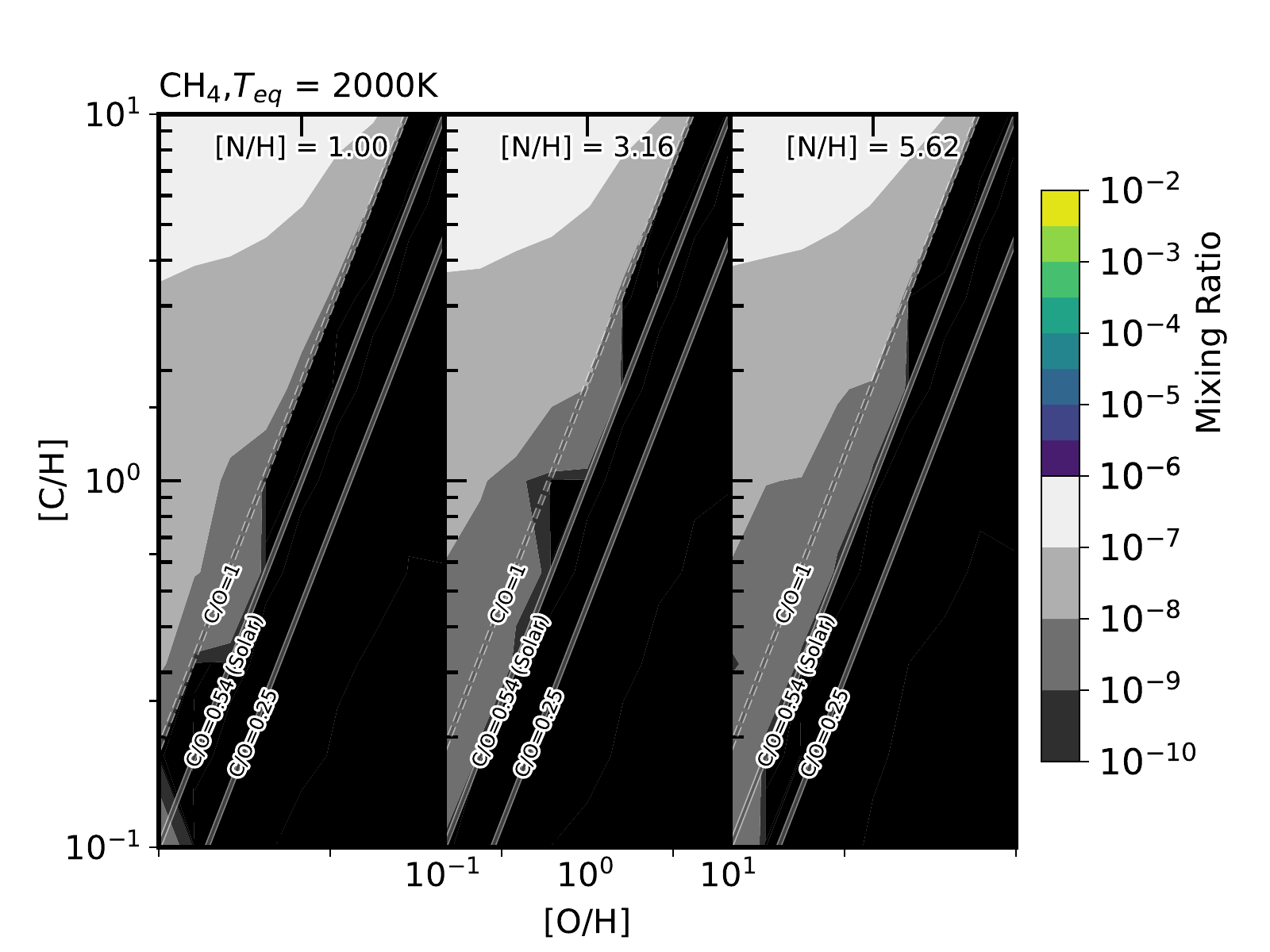}
    \caption[Methane abundance across the parameter space]{The abundance of \ce{CH4} for two hot Jupiters, one with a $1000\,\mathrm{K}$ equilibrium temperature (top) and one with a $2000\,\mathrm{K}$ equilibrium temperature (bottom). The variation in abundance of \ce{CH4} is shown against the atmospheric C/H and O/H ratios normalised to solar values. The C/O ratios of 0.25, 0.54 (solar) and 1 are shown on each plot to assist visualisation. Additionally, we consider three atmospheric N/H ratios normalised to solar; 1, 3.2 and 5.6. These are shown from left to right on the figures above.}
     \label{fig:CH4}
\end{figure}

The variation in the methane abundance across our parameter space is shown in Figure \ref{fig:CH4}. We find \ce{CH4} has a positive dependence upon the C/H ratio and a negative dependence upon the O/H ratio for the two hot Jupiters we model. This is due to \ce{CO} sequestering a greater fraction of the carbon as the amount of available O in the atmosphere increases. Unlike \ce{H2O} and \ce{CO}, methane's abundance also has a strong temperature dependence. At $1000\,\mathrm{K}$, \ce{CH4} has a maximal abundance of $10^{-2}$ in the high C/H, low O/H regime, with a minimal abundance of $10^{-8}$ in the low C/H, high O/H regime. By comparison, at $2000\,\mathrm{K}$, the maximum and minimum abundance of methane is $10^{-6}$ and $10^{-16}$ respectively. Methane's abundance is approximately unchanging along lines of constant C/O ratio, making it an excellent check to confirm the values of the C/O first expected by examining \ce{CO} and \ce{H2O}. However, it will likely be impossible to detect \ce{CH4} on the hotter hot Jupiters; in our $2000\,\mathrm{K}$ model, for a C/O < 1, methane's abundance is typically below $10^{-9}$. This is far below the estimates we have for the detectable limits of molecules. The strong dependence of the \ce{CH4} abundance on temperature does make it a good proxy for the temperature, but a poor tool for examining the metalicity of the very hot Jupiters.

\subsection{\texorpdfstring{\ce{CO2}}{CO2} abundance}
\begin{figure}
        \centering
        \includegraphics[width=0.475\textwidth]{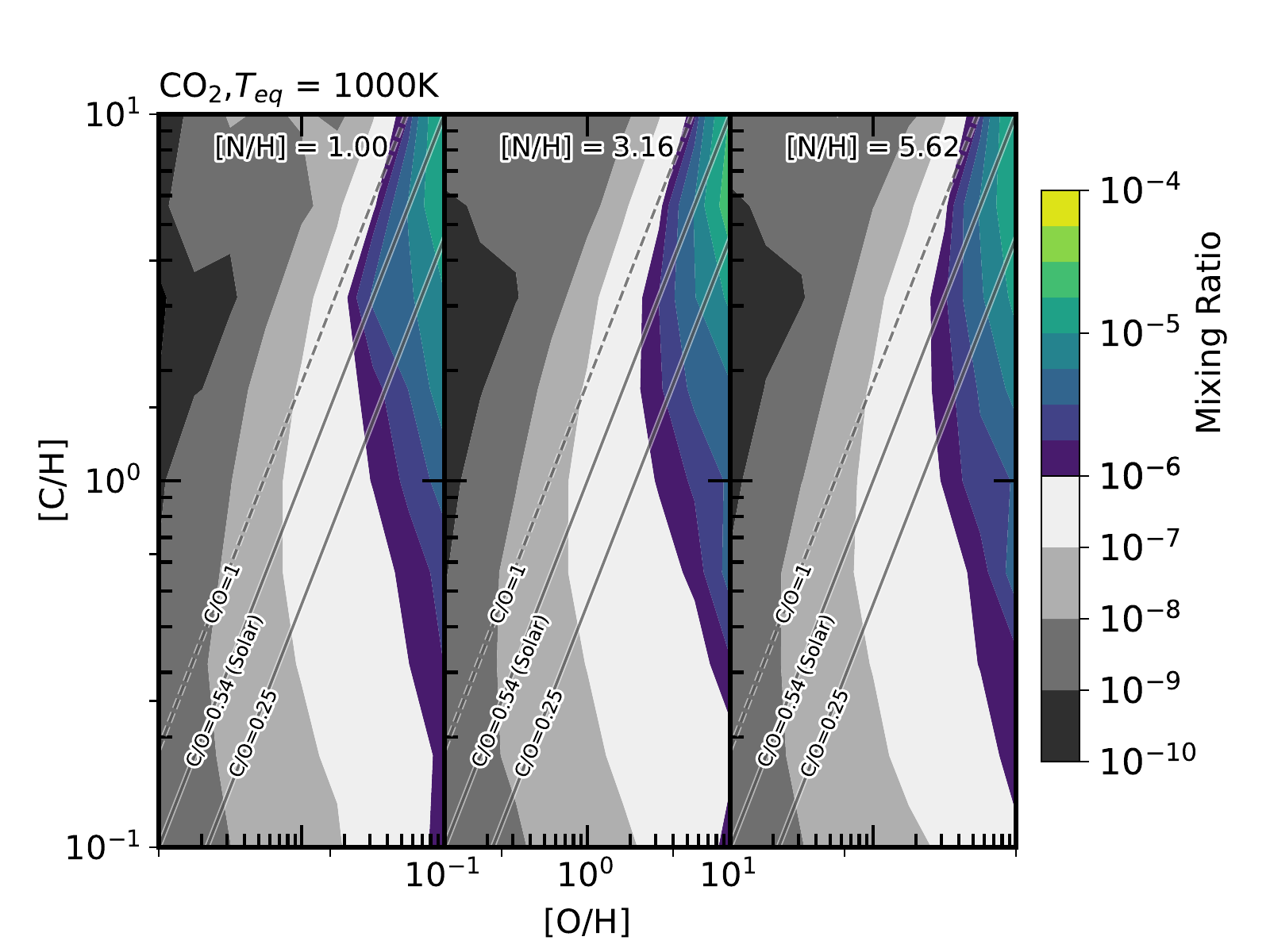}
        \includegraphics[width=0.475\textwidth]{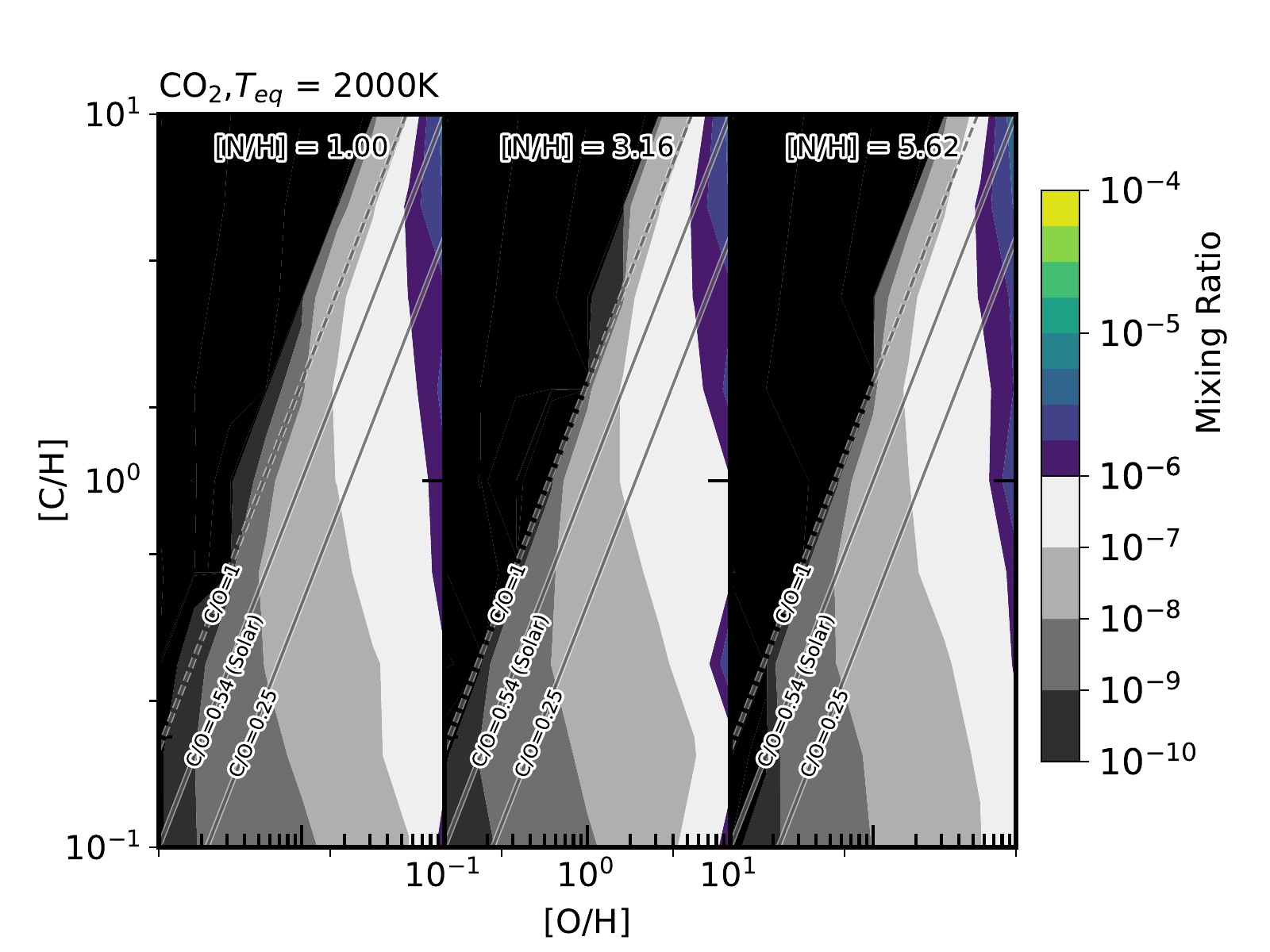}
    \caption[\ce{CO2} abundance across the parameter space]{The abundance of \ce{CO2} for two hot Jupiters, one with a $1000\,\mathrm{K}$ equilibrium temperature (top) and one with a $2000\,\mathrm{K}$ equilibrium temperature (bottom). The variation in abundance of \ce{CO2} is shown against the atmospheric C/H and O/H ratios normalised to solar values. The C/O ratios of 0.25, 0.54 (solar) and 1 are shown on each plot to assist visualisation. Additionally, we consider three atmospheric N/H ratios normalised to solar; 1, 3.2 and 5.6. These are shown from left to right on the figures above.}
     \label{fig:CO2}
\end{figure}

We show the range of \ce{CO2} abundances in Figure \ref{fig:CO2}. We find that \ce{CO2} is very strongly dependent on the O/H ratio, but only weakly dependent on the C/H ratio in our models. However, there is a significant decrease in \ce{CO2} abundance when the C/O ratio crosses from less than 1 to greater than 1. This is because, for C/O > 1, the majority of the oxygen is sequestered in the form of \ce{CO}, leaving little available for other oxygen bearing species. The range of \ce{CO2} abundances in both temperature cases are similar for C/O ratios less than 1, ranging between $10^{-9}$ and $10^{-4}$. Though for C/O > 1, \ce{CO2} follows a similar pattern to \ce{H2O}, for $T_{eq}$ = $2000\,\mathrm{K}$ there is a rapid decline in abundance as the C/O ratio increases, while for $T_{eq}$ = $1000\,\mathrm{K}$ the decline is much slower. \ce{CO2} is a poor diagnostic tool by itself, with a large range of both C/H and O/H ratios corresponding to a single abundance.

\subsection{\texorpdfstring{\ce{HCN}}{HCN} abundance}
\begin{figure}
        \centering
        \includegraphics[width=0.475\textwidth]{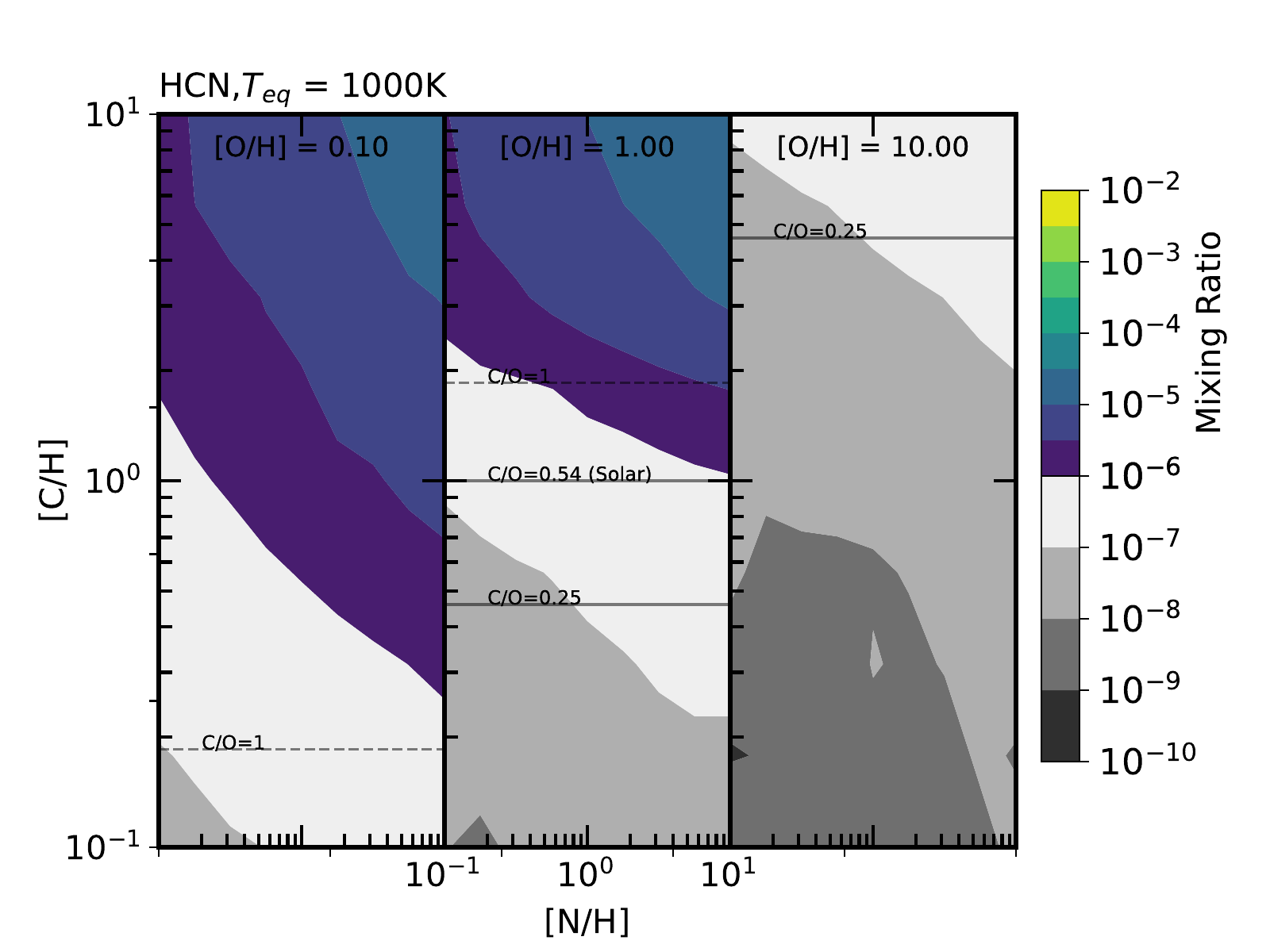}
        \includegraphics[width=0.475\textwidth]{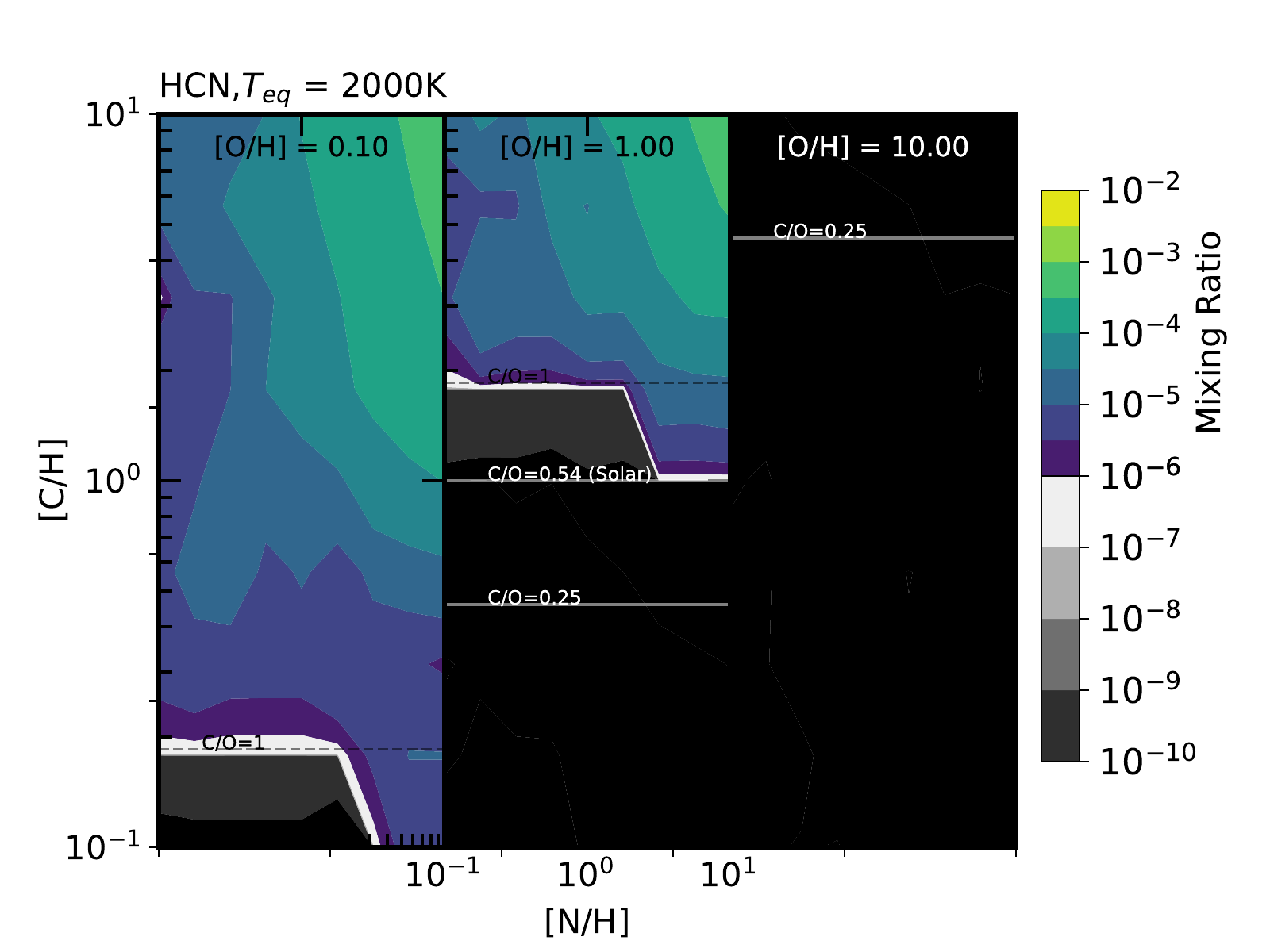}
    \caption[\ce{HCN} abundance across the parameter space]{The abundance of \ce{HCN} for two hot Jupiters, one with a $1000\,\mathrm{K}$ equilibrium temperature (top) and one with a $2000\,\mathrm{K}$ equilibrium temperature (bottom). The variation in abundance of \ce{HCN} is shown against the atmospheric C/H and N/H ratios normalised to solar values. The C/O ratios of 0.25, 0.54 (solar) and 1 are shown on each plot to assist visualisation. Additionally, we consider three atmospheric O/H ratios normalised to solar; 0.1, 1 and 10. These are shown from left to right on the figures above.}
     \label{fig:HCN}
\end{figure}

In Figure \ref{fig:HCN} we present the range of \ce{HCN} abundances across our parameter space. Unlike in the previous figures in this section, we have replaced the O/H ratio on the x-axis with the N/H ratio, with three values for the O/H ratio chosen to examine. As expected, \ce{HCN} is strongly dependent on both the C/H and N/H ratios at the temperatues we model. While some dependence on the O/H ratio is also observed, this is merely a consequence of the O/H ratio affecting the overall C/O ratio.

We see a rapid increase in the abundance of \ce{HCN} in the atmosphere for C/O ratios greater than 1. This is because the most abundant carbon carrier, \ce{CO}, is no longer limited by the available carbon in the atmosphere, but by the available oxygen at these ratios. Thus, additional \ce{HCN} can form due to the excess carbon available. Both planetary temperatures modelled have similar maximum abundances for \ce{HCN}, at around $10^{-4}$, at 10 [N/H] and C/O $\gg$ 1. For C/O less than 1, we find that for lower temperatures we expect \ce{HCN} abundances around $10^{-7}$. For the hotter temperatures the \ce{HCN} abundance is much lower, around $10^{-11}$. The \ce{HCN} abundance can function as a way of tracing the N/H ratio, however, for C/O < 1, highly sensitive measurements will be needed to detect \ce{HCN}'s predicted low abundances.

\subsection{\texorpdfstring{\ce{NH3}}{NH3} abundance}

\begin{figure}
        \centering
        \includegraphics[width=0.475\textwidth]{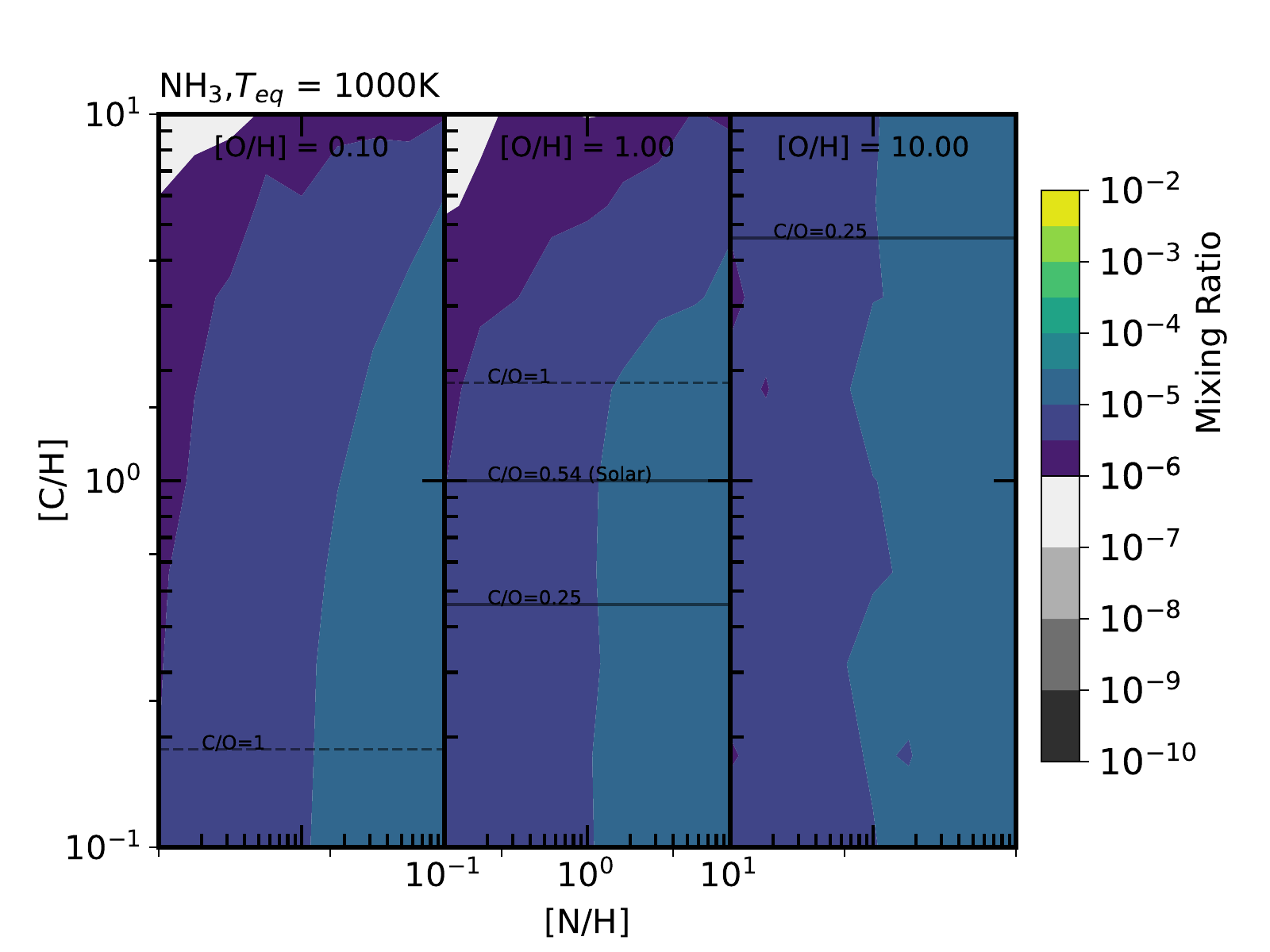}
        \includegraphics[width=0.475\textwidth]{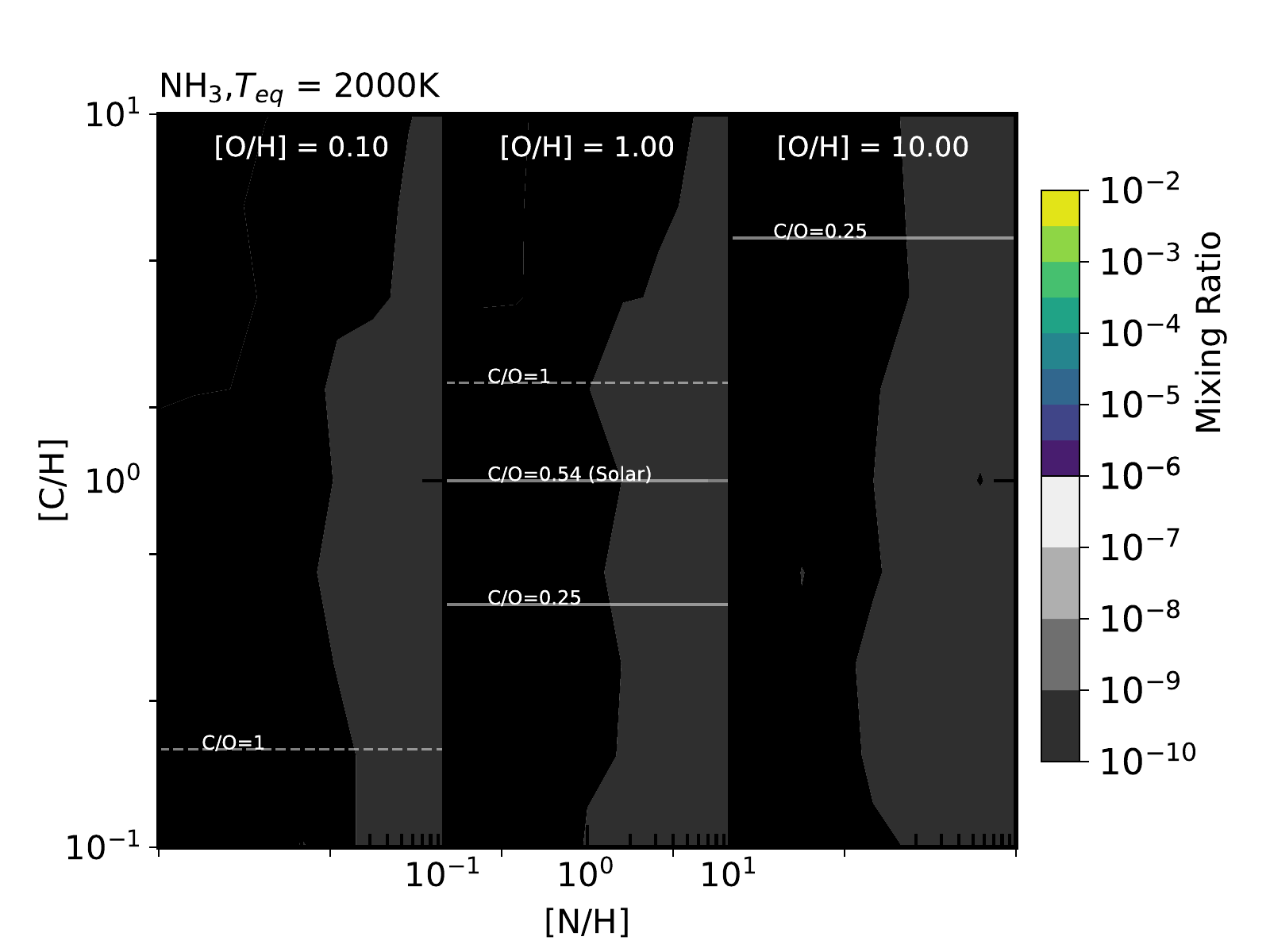}
    \caption[Ammonia abundance across the parameter space]{The abundance of \ce{NH3} for two hot Jupiters, one with a $1000\,\mathrm{K}$ equilibrium temperature (top) and one with a $2000\,\mathrm{K}$ equilibrium temperature (bottom). The variation in abundance of \ce{NH3} is shown against the atmospheric C/H and N/H ratios normalised to solar values. The C/O ratios of 0.25, 0.54 (solar) and 1 are shown on each plot to assist visualisation. Additionally, we consider three atmospheric O/H ratios normalised to solar; 0.1, 1 and 10. These are shown from left to right on the figures above.}
     \label{fig:H3N}
\end{figure}

In Figure \ref{fig:H3N} we present how the \ce{NH3} abundance varies across our parameter space. As expected, \ce{NH3} is dependent on the N/H ratio. We find that \ce{NH3} is also dependent on the C/H and O/H ratio to a small extent, however this is mainly a function of the global C/O ratio, not the individual ratios themselves. Where the C/O ratio is greater than 1, we see large abundances of \ce{HCN}, limiting the amount of available N to form \ce{NH3}. We also see large differences in ammonia abundance between the two temperature models. At lower temperatures, with C/O < 1, the ammonia abundance varies between $10^{-5}$ and $10^{-4}$. While in our higher temperature model we find approximately five orders of magnitude less ammonia in the atmosphere, between $10^{-10}$ and $10^{-9}$ for C/O < 1. The abundance of \ce{NH3} is a useful tool to determine the N/H ratio for cooler planets, as long as sufficiently sensitive measurements to detect it in the atmosphere can be made.

\section{Comparison with Formation and Migration Models} 
\label{secF:Section4}

In this section we use our results from the previous section in conjunction with the formation models of \cite{Madhu2014,Booth2017,Turrini2021}. We compare the metalicity ranges these works predict for different formation and migration models to the parameter figures in the previous section. This gives us a abundance range, for each of the six molecules investigated previously, for each formation and migration model. Thus, we can define what our model predicts the atmospheric composition of a hot Jupiter will be based upon where the planet formed and how it migrated. One point of consideration is that we do not include cloud formation in our models. Previous works \citep[e.g.][]{Helling2021} show that clouds can deplete oxygen in the gas reservoir of the atmosphere, thus effectively increasing the C/O ratio.

\subsection{Planetary Carbon and Oxygen abundances due to migration}

 From the work of \cite{Madhu2014}, for a hot Jupiter formed via core accretion with subsequent disk migration to its current location, we expect [C/H] ratios between 1 and 5, and [O/H] ratios between 1 and 10, and that the C/O ratio is always less than solar. In Table \ref{tab:Madhu1} we show the range of expected chemical abundances of the six molecules we investigated in the previous section for this elemental parameter space. 

\begin{table}
    \centering
    \caption[Expected abundance ranges for hot Jupiters formed by core accretion that migrate within the disk]{Expected abundance ranges for hot Jupiters formed by core accretion that migrate within the disk}
    \begin{tabular}{p{2cm} c c}
    \hline

        Molecule   & $\textrm{log}({X_{\textrm{Max}}, X_{\textrm{Min}}})$, $1000\,\mathrm{K}$  & $\textrm{log}({X_{\textrm{Max}}, X_{\textrm{Min}}})$, $2000\,\mathrm{K}$ \\ \hline
        \ce{H2O}   & (-2, -3.3)           &  (-2, -4) \\
        \ce{CO}    & (-2.2, -3.2)         &  (-2, -3.3)   \\
        \ce{CH4}   & (-5, -6)             & ( -13, -14)  \\
        \ce{CO2}   & (-4, -7)             & (-5, -7)  \\
        \ce{HCN}   & (-7, -8)             & (-10, -12)  \\
        \ce{NH3}   & (-4.3, -5.3)         & (-9.3, -10.3) \\
        \hline

    \end{tabular}
 The expected mixing ratios for six molecules in the atmospheres of two hot Jupiters, one with $T_{eq}=1000\,\mathrm{K}$ and the other with $T_{eq}=2000\,\mathrm{K}$, for a gas giant formed by core accretion that migrated within the disk, or a gas giant formed by gravitational instability between the \ce{CO2} and \ce{CO} snowline that migrated disk-free, from the work of \protect\cite{Madhu2014}.
    \label{tab:Madhu1}
\end{table}

We split planets formed by core accretion that then underwent disk-free migration into two groups: Those that formed beyond the \ce{CO2} snowline, with sub-solar C/H and O/H and those that formed within the \ce{CO2} snowline, with super-solar C/H and O/H. From \cite{Madhu2014}, planets that formed within the \ce{CO2} snowline, have allowed metalicities along the straight line between [C/H] and [O/H] equal to 1, and [C/H] = 2 and [O/H] = 4, and planets that formed beyond the \ce{CO2} snowline have allowed metalicities along the straight line between [C/H] and [O/H] equal to 1, and [C/H] = 0.6 and [O/H] = 0.4. \cite{Madhu2014} did not consider the N/H in their models, and so we choose to use N/H = 1 for the comparisons here. We present the expected abundance for these planets in Table \ref{tab:Madhu2}.

\begin{table}
    \centering
    \caption[Expected abundance ranges for hot Jupiters formed by core accretion that migrate disk-free]{Expected abundance ranges for hot Jupiters formed by core accretion that migrate disk-free}
    \begin{tabular}{c c c}

    \hline
        
        Molecule   &  $1000\,\mathrm{K}$  & $2000\,\mathrm{K}$ \\ \hline
         Within the \ce{CO2} snowline &$\textrm{log}({X_{\textrm{Max}}, X_{\textrm{Min}}})$ &$\textrm{log}({X_{\textrm{Max}}, X_{\textrm{Min}}})$\\ \hline
        \ce{H2O}   & (-2.3, -3.3)    &  (-2, -4)  \\
        \ce{CO}    & (-3, -3.2)    &   (-2, -3.3)    \\
        \ce{CH4}   &(-4, -6)     & (-13, -15)   \\
        \ce{CO2}   &(-5, -7)     & (-6, -8)   \\
        \ce{HCN}   & (-7, -8)    & (-11, -12)   \\
        \ce{NH3}   & (-4.3, -5.3)    & (-9.3, -10.3)  \\
    \hline
    Beyond the \ce{CO2} snowline\\ \hline
        \ce{H2O}   & (-3, -4)   &  (-3, -5) \\
        \ce{CO}    & (-3.3, -3.5)   &  (-3, -3.3)   \\
        \ce{CH4}   &(-4, -5)    & (-8, -11)  \\
        \ce{CO2}   &(-6, -8)    & (-7, -9)  \\
        \ce{HCN}   & (-6, -7)   & (-9, -11)  \\
        \ce{NH3}   & (-4.3, -5.3)    & (-9.3, -10.3) \\
    \hline
    \end{tabular}
    The expected mixing ratios for six molecules in the atmospheres of two hot Jupiters, one with $T_{eq}=1000\,\mathrm{K}$ and the other with $T_{eq}=2000\,\mathrm{K}$, for a gas giant formed by core accretion that underwent disk-free migration, from the work of \protect\cite{Madhu2014}. We split the formation location of the planet into two groups: Within the \ce{CO2} snowline and beyond the \ce{CO2} snowline.
    \label{tab:Madhu2}
\end{table}

Lastly from the work of \cite{Madhu2014}, we consider planets formed via gravitational instability that then migrate inwards disk-free. There are again two regions to consider for this formation-migration mechanism: Those planets that formed within the CO snowline and those that formed outside the CO snowline. The composition of those planets that formed by gravitational instability within the CO snowline is similar to those formed by core accretion with in-disk migration, however the potential metalicity also extends into sub-solar metalicity with super-solar C/O ratio. Those that formed beyond the CO snowline have a near solar C/O ratio, but with any metalicity within our parameter space, producing a wide range of possible abundances. We present the expected abundance ranges for both of these cases in Table \ref{tab:Madhu3}.

\begin{table}
    \centering
    \caption[Expected abundance ranges for hot Jupiters formed by gravitational instability that migrate disk-free]{Expected abundance ranges for hot Jupiters formed by gravitational instability that migrate disk-free}
    \begin{tabular}{c c c}
    \hline
        
        Molecule   &  $1000\,\mathrm{K}$  & $2000\,\mathrm{K}$ \\ \hline
        Within the \ce{CO} snowline   & $\textrm{log}({X_{\textrm{Max}}, X_{\textrm{Min}}})$  & $\textrm{log}({X_{\textrm{Max}}, X_{\textrm{Min}}})$\\ \hline
        \ce{H2O}   & (-2, -4)           &  (-2, -4) \\
        \ce{CO}    & (-2.2, -3.5)         &  (-2, -4)   \\
        \ce{CH4}   & (-3.3, -6)             & ( -8.5, -14)  \\
        \ce{CO2}   & (-4, -7.5)             & (-5, -10)  \\
        \ce{HCN}   & (-6, -8)             & (-6, -12)  \\
        \ce{NH3}   & (-4.3, -5.3)         & (-9.3, -10.3) \\   
        
        \hline
        
        Beyond the \ce{CO} snowline\\\hline
        \ce{H2O}   & (-2, -5)   &  (-2, -5) \\
        \ce{CO}    & (-2.5, -5)   &  (-2, -4.5)   \\
        \ce{CH4}   &(-4, -5)    & (-12, -13)  \\
        \ce{CO2}   &(-4, -9)    & (-5, -10)  \\
        \ce{HCN}   & (-6.3, -7)   & (-10, -11)  \\
        \ce{NH3}   & (-4.3, -5.3)    & (-9.3, -10.3) \\
        \hline

    \end{tabular}
    The expected mixing ratios for six molecules in the atmospheres of two hot Jupiters, one with $T_{eq}=1000\,\mathrm{K}$ and the other with $T_{eq}=2000\,\mathrm{K}$, for a gas giant formed by gravitational instability and underwent disk-free migration from the work of \protect\cite{Madhu2014}. We split this formation mechanism for those planets formed outside of the CO snowline and those that formed within the CO snowline but beyond the \ce{CO2} snowline.
    \label{tab:Madhu3}
\end{table}

We summarise the resulting atmospheric chemistry at $10^{-3}\,\textrm{bar}$ for our $1000\,\mathrm{K}$ and $2000\,\mathrm{K}$ hot Jupiters following formation and migration from these five different scenarios in Figure \ref{fig:Abundcomp}. Overall what we see is significant overlap between formation location and migration types investigated here. This was expected based upon the overlapping metalicities that we drew from \cite{Madhu2014}. We can still draw some important insights from this data though. 

\textit{\ce{H2O}}: We find that high \ce{H2O} abundances in our models, between $10^{-2}$ and $10^{-3}$, can be found for a hot Jupiter that formed at any location, and is independent of the equilibrium temperature of the hot Jupiter. However, very low \ce{H2O} abundances were unique to planets that had formed further out, at least beyond the \ce{CO2} snowline for both core accretion and gravitational instability formation models. This is because these formation locations resulted in either low O/H ratio or a high C/O ratio, both of which cause low \ce{H2O} abundances.

\textit{\ce{CO}}: Similar to \ce{H2O}, our model predicts high \ce{CO} abundances in both cases are able to occur regardless of where the hot Jupiter formed. However, we once again find that very low \ce{CO} abundances are only expected for planets that formed further out, either by core accretion beyond the \ce{CO2} snowline or by gravitational instability beyond the \ce{CO} snowline for our 1000K model, or any location of gravitational instability for our 2000K model. This is due to low C/H and O/H ratios, irrespective of the C/O ratio.

\textit{\ce{CH4}}: For our 2000K temperature hot Jupiter, we never expect \ce{CH4} to be detectable, with abundances always far below $10^{-6}$. For our cooler 1000K hot Jupiter, methane abundances can be high, up to $10^{-4}$ for planets formed by core accretion with disk-free migration or gravitational instability beyond the CO snowline. We expect to see the highest \ce{CH4} levels in the case of the 1000K hot Jupiter that formed between the \ce{CO2} and \ce{CO} snowlines, possibly approaching an abundance of $10^{-3}$. This is because this is the model with the highest C/O ratio, on which the \ce{CH4} ratio strongly depends. For our 1000K hot Jupiter, we only find methane abundances below $10^{-5}$ for planets that formed close in by core accretion, either within the \ce{CO2} snowline and then migrated disk-free or those that underwent in-disk migration, or formed within the \ce{CO} snowline by gravitational instability. It is possible then that detections of low levels of \ce{CH4} could also function in helping determine where a hot Jupiter formed.

\textit{\ce{CO2}}: For core accretion, our models find higher levels of \ce{CO2} are associated with a planet that formed further in, with both the maximum and minimum levels of \ce{CO2} increasing by up to two orders of magnitude as we move from the more distant formation models to the closer ones. However, planets that formed by gravitational instability, can contain a range of \ce{CO2} levels that encompasses the entire range of all the other models, requiring other features to break the degeneracy here. 

\textit{\ce{HCN}}: For our 2000K temperature hot Jupiter, we predict \ce{HCN} abundances below $10^{-8}$, and thus very unlikely to be detectable, for every model except one: We find that our 2000K hot Jupiter formed by gravitational instability between the \ce{CO2} and \ce{CO} snowlines can have \ce{HCN} abundances as high as $10^{-6}$. In the atmosphere of our 1000K hot Jupiter, we only find high levels of \ce{HCN} in the model of core accretion and then disk-free migration for planets that formed beyond the \ce{CO2} snowline and gravitational instability within the \ce{CO} snowline. This is because these are the only models with a significantly super-solar C/O ratio, which favours the production of \ce{HCN}. 

\textit{\ce{NH3}}: Ammonia levels cover the same range in every model, with no expected metalicity to be extreme enough to find significantly different \ce{NH3} levels. However, for our 2000K temperature hot Jupiter, the abundance of \ce{NH3} in their atmospheres is expected to be below $10^{-8}$, and thus not detectable. For the 1000K hot Jupiter, \ce{NH3} is expected to be detectable at around $10^{-5}$.

\subsection{Chemical Enrichment of hot Jupiters}
\begin{figure*}
        \centering
        \includegraphics[width=1\textwidth]{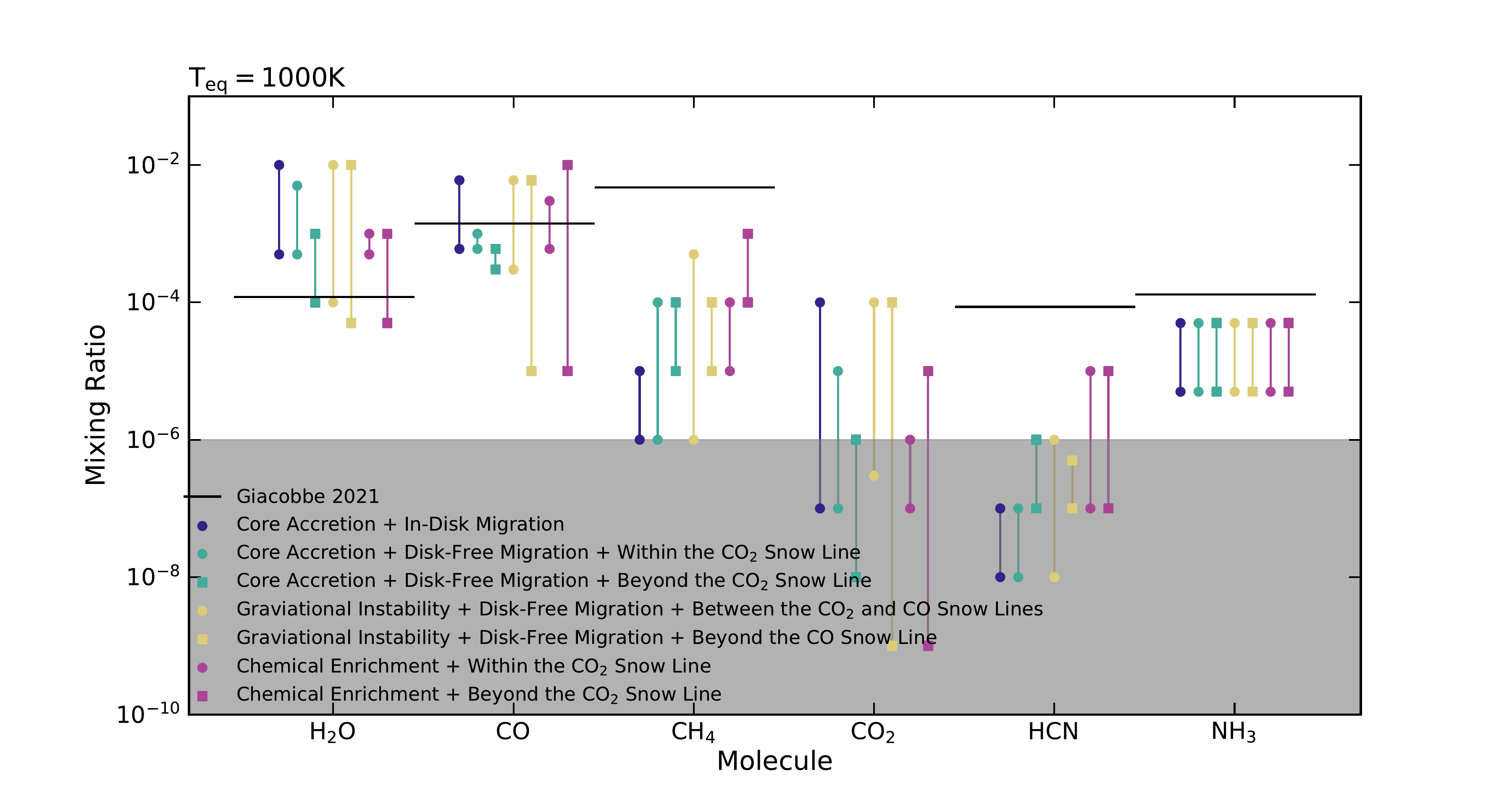}
        \includegraphics[width=1\textwidth]{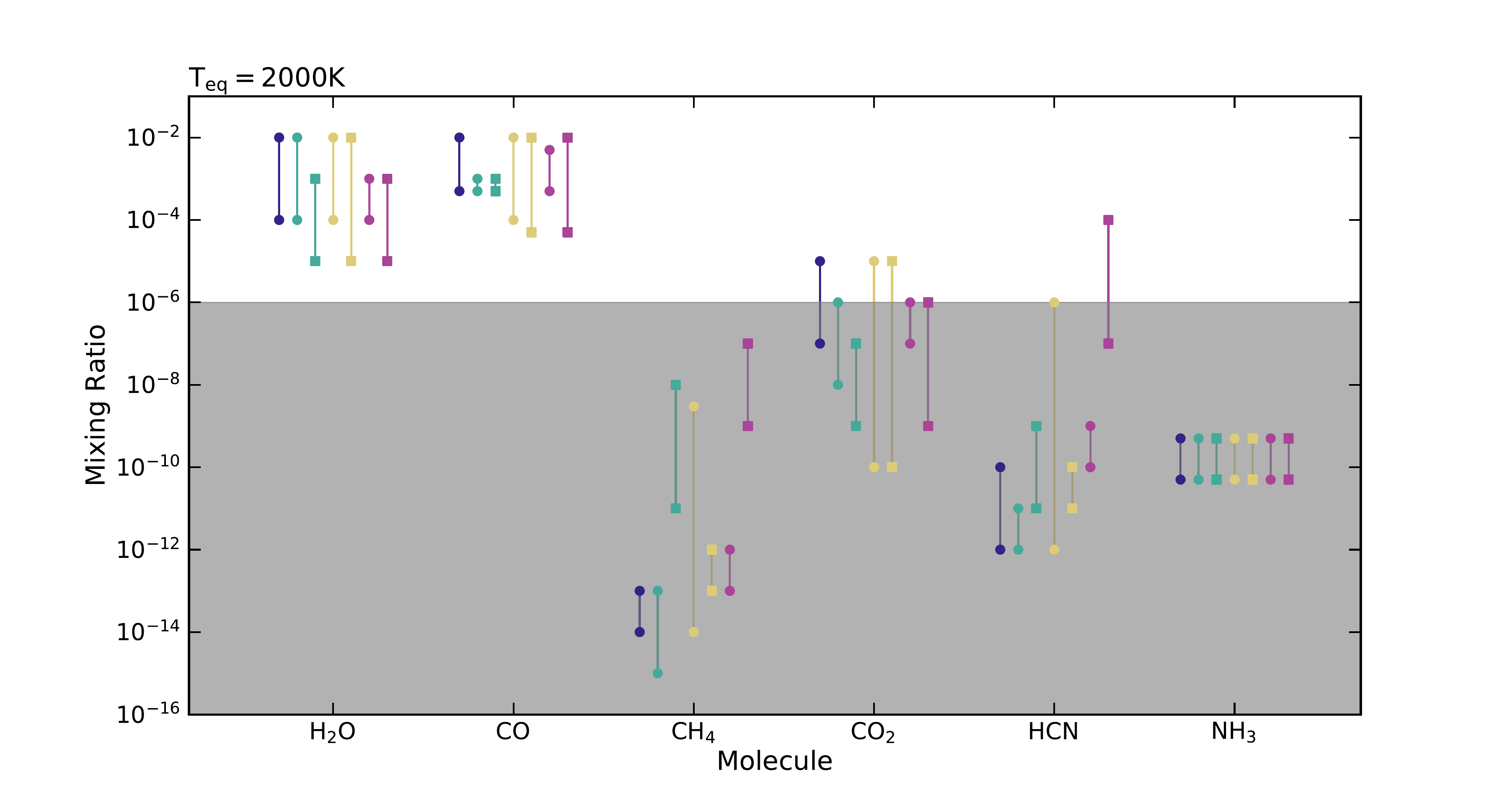}
    \caption[Predicted abundance range from five formation scenarios]{A comparison of the abundances of six molecules that our models predict, based upon the metalicities resulting from the five different formation and migration mechanisms discussed in \protect\cite{Madhu2014} and two different formation locations from \protect\cite{Booth2017}, for the two different planet temperatures we investigate. The red, blue and green ranges correspond to work from \protect\cite{Madhu2014} while the purple ranges correspond to work from \protect\cite{Booth2017}. The range of abundances we see for each species for each formation model is due to the spread of metallicity possible from those models. The grey zone below $10^{-6}$ marks the region in which abundances are expected to be too low to be detectable. The black horizontal lines indicated representative abundances for key molecules detected in \citep{Giacobbe2021}.}
     \label{fig:Abundcomp}
\end{figure*}

The work of \cite{Booth2017} looks at the chemical enrichment by pebble drift of gas giants formed by core accretion. For planets formed within the \ce{CO2} snowline this results in 2 < [C/H] < 3 and [O/H] = 2 and for planets formed beyond the \ce{CO2} snowline this can result in any metalicity within our parameter space, but with a C/O = 1. The abundances our model predicts for this formation and migration scenario are in Table \ref{tab:Booth1}. We summarise the resulting atmospheric chemistry in $1000\,\mathrm{K}$ and $2000\,\mathrm{K}$ hot Jupiters following formation and migration from these two different scenarios in Figure \ref{fig:Abundcomp}.

\begin{table}
    \centering
     \caption[Expected abundance ranges for hot Jupiters that became chemically enriched via pebble-drift]{Expected abundance ranges for hot Jupiters that became chemically enriched via pebble-drift}
    \begin{tabular}{c c c}
    
    \hline
        
       Molecule   &  $1000\,\mathrm{K}$  & $2000\,\mathrm{K}$ \\ \hline
        Within the \ce{CO2} snowline & $\textrm{log}({X_{\textrm{Max}}, X_{\textrm{Min}}})$ & $\textrm{log}({X_{\textrm{Max}}, X_{\textrm{Min}}})$\\
        \hline
        \ce{H2O}   & (-3, -4.3)   &  (-3, -4) \\
        \ce{CO}    & (-2.5, -3.2)   &  (-2.3, -3.3)   \\
        \ce{CH4}   &(-4, -5)    & (-12, -13)  \\
        \ce{CO2}   &(-6, -7)    & (-6, -7)  \\
        \ce{HCN}   & (-5, -7)   & (-9, -10)  \\
        \ce{NH3}   & (-4.3, -5.3)    & (-9.3, -10.3) \\
    \hline
    
        Beyond the \ce{CO2} snowline\\
        \hline
        \ce{H2O}   & (-3, -4.3)   &  (-3, -5) \\
        \ce{CO}    & (-2, -5)   &  (-2, -4.3)   \\
        \ce{CH4}   &(-3, -4)    & (-7, -9)  \\
        \ce{CO2}   &(-5, -9)    & (-6, -9)  \\
        \ce{HCN}   & (-5, -7)   & (-4, -7) \\
        \ce{NH3}   & (-4.3, -5.3)    & (-9.3, -10.3) \\
    \hline
    \end{tabular}
   The expected mixing raitos for six molecules in the atmospheres of two hot Jupiters, one with $T_{eq}=1000\,\mathrm{K}$ and the other with $T_{eq}=2000\,\mathrm{K}$, using the models of chemical enrichment by \protect\cite{Booth2017}. We split the formation location of the planet into two groups: Within the \ce{CO2} snowline and beyond the \ce{CO2} snowline.
    \label{tab:Booth1}
\end{table}

Overall what we see is that only \ce{CH4} and \ce{HCN} in chemically enriched planets' atmospheres can stand out as tracers of these planets' pasts. Regardless of temperature, \ce{H2O}, \ce{CO}, \ce{CO2} and \ce{NH3} all lie within the ranges of expected abundances discussed in the previous section. For only the 2000K hot Jupiters, do we find that chemical enrichment of planets formed beyond the \ce{CO2} snowline leads to large elevations in the expected abundance of \ce{HCN}. Also, our predicted \ce{CH4} abundance in planets that have been chemically enriched by pebble drift having formed beyond the \ce{CO2} snowline, are at least half an order of magnitude above values expected by other formation models, making it the best contender for a chemical enrichment tracer. 

There are a number of tracers to distinguish where a planet that was known to have been chemically enriched was formed. Both \ce{H2O} and \ce{CO} abundances in enriched planets that formed beyond the \ce{CO2} snowline are expected to be able to reach significantly lower values than in a planet formed within the \ce{CO2} snowline. In hotter hot Jupiters, \ce{HCN} can help distinguish two formation areas, however in cooler hot Jupiters there is a full overlap in the expected abundances of \ce{HCN} from both locations. \ce{CH4} likely works as the best tracer however. Regardless of temperature, there is no overlap in expected abundance of \ce{CH4} between the two formation areas, with significantly higher \ce{CH4} always being expected for a planet that formed beyond the \ce{CO2} snowline.

\subsection{Tracers of nitrogen chemistry}
\cite{Turrini2021} select six formation locations in conjunction with a core accretion and in-disk model to produce the final metalicity of a hot Jupiter. These values are shown in Table \ref{tab:Turrinibulk}. We tabulate the molecular abundance our model predicts for each planetary formation location in Table \ref{tab:TurriniAbun}. 

\begin{table}
    \centering
    \caption[Expected abundances for a hot Jupiters that formed at different locations within the disk]{Expected abundances for a hot Jupiters that formed at different locations within the disk}
    \begin{tabular}{c c c c c c c}

    \hline
        
        Molecule   & 5AU  & 12AU & 19AU & 50AU & 100AU & 130AU\\ \hline
       $1000\,\mathrm{K}$ &$\textrm{log}({X})$&$\textrm{log}({X})$&$\textrm{log}({X})$&$\textrm{log}({X})$&$\textrm{log}({X})$& $\textrm{log}({X})$\\ \hline
        \ce{H2O}   & -3.3   &  -3.2 & -3.1 & -3.0 & -2.7 & -2.5\\
        \ce{CO}    & -3.4   &  -3.2 & -3.1 & -2.7 & -2.7 & -2.5   \\
        \ce{CH4}   & -4.4   &  -4.5 & -4.2 & -4.7 & -5 & -5.1  \\
        \ce{CO2}   & -6.7   &  -6.5 & -6.1 & -5.5 & -5.2 & -5.0 \\
        \ce{HCN}   & -6.5   &  -6.2 & -6.3& -6.3 & -6.1 & -6.0  \\
        \ce{NH3}   & -5.1   &  -5.1 & -5.0 & -5.0 & -4.7 & -4.7 \\
    \hline
    $2000\,\mathrm{K}$\\ \hline
    
        \ce{H2O}   & -3.3   &  -3.3 & -3.0 & -2.3 & -2.3 & -2.3 \\
        \ce{CO}    & -3.3   &  -3.3 & -3.0 & -3.0 & -2.5 & -2.3   \\
        \ce{CH4}   & -12   &  -12 & -12 & -12 & -12 & -12  \\
        \ce{CO2}   & -7.3   &  -7.0 & -7.0 & -6.3 & -6.0 & -5.3 \\
        \ce{HCN}   & -10.3   &  -10.3 & -10.3 & -10.2 & -10.1 & -10  \\
        \ce{NH3}   & -10.2   &  -10.1 & -10 & -10 & -9.7 & -9.5 \\
    \hline
    \end{tabular}
    The expected mixing ratios for six molecules in the atmospheres of two hot Jupiters, one with $T_{eq}=1000\,\mathrm{K}$ and the other with $T_{eq}=2000\,\mathrm{K}$, using the models \protect\cite{Turrini2021}. We split the table into columns of formation location of the planet, using the metalicity of these planets shown in Table \ref{tab:Turrinibulk} in conjunction with our models to produce the abundances shown in this table.
    \label{tab:TurriniAbun}
\end{table}

\begin{figure}
        \centering
        \includegraphics[width=0.475\textwidth]{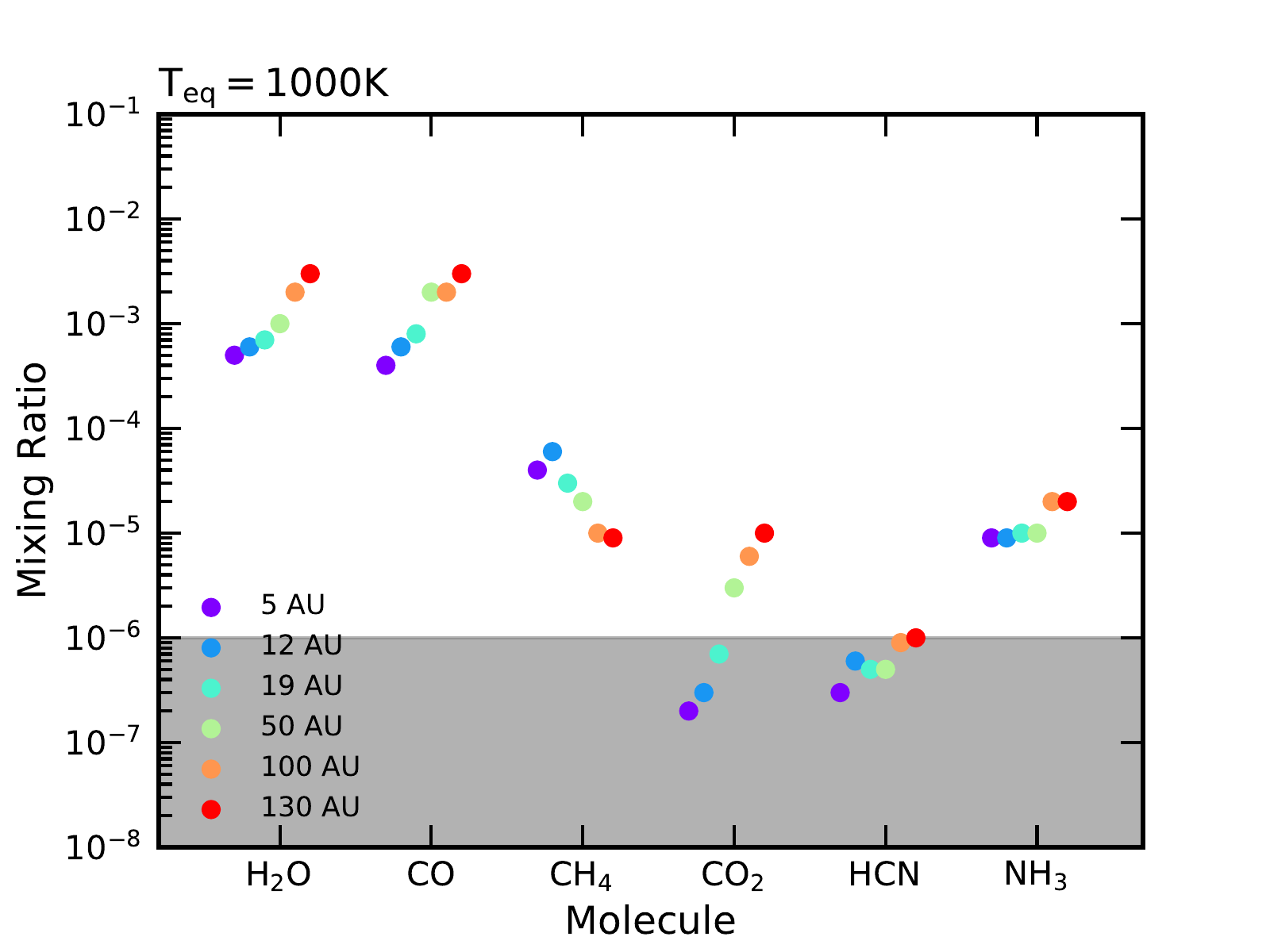}
        \includegraphics[width=0.475\textwidth]{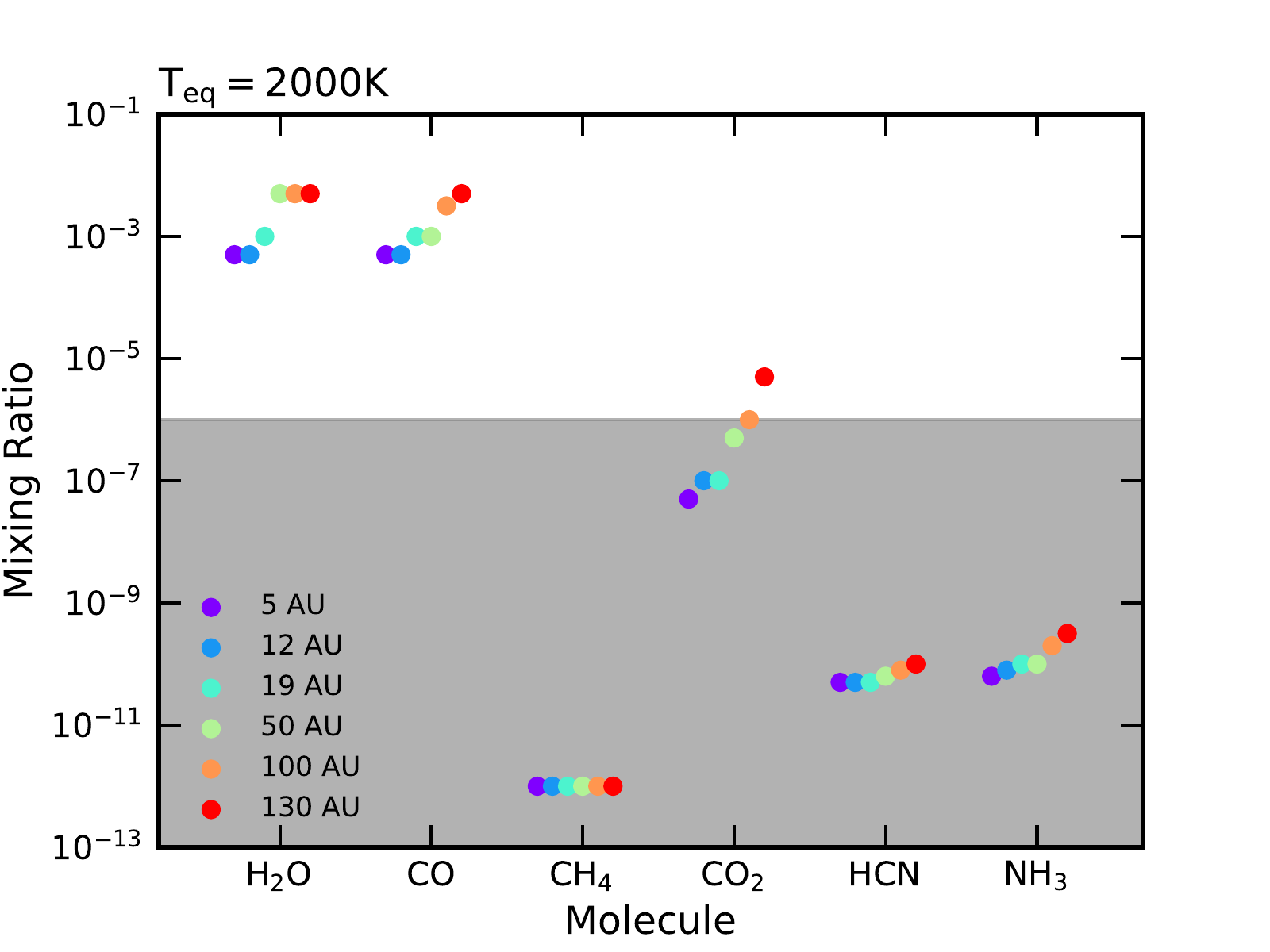}
    \caption[Predicted abundances from a formation scenario considering the N/H ratio]{A comparison of the abundances of six molecules that our models predict for the different formation locations presented in \protect\cite{Turrini2021}, for hot Jupiters at two equilibrium temperatures. The grey zone below $10^{-6}$ marks the region in which abundances are expected to be too low to be detectable. The black lines for $T_eq$ = 1000K correspond to the values taken from \protect\cite{Giacobbe2021}, for comparisons made in Section \ref{secF:Section5}}.
     \label{fig:T_Abundcomp}
\end{figure}

As expected, the monotonic increase in the metallicity of hot Jupiters as formation distance increases, translates to a similar monotonic increase for each of the six molecules we examine. This is shown in Figure \ref{fig:T_Abundcomp}. Notably, the molecules \ce{H2O}, \ce{CO}, \ce{CH4} and \ce{CO2} all change by approximately an order of magnitude across the range of formation locations. This should be sufficient for detections of these molecules in a planet's atmosphere to link quite accurately to the planet's C/O ratio and where it formed. However, \ce{HCN} and \ce{NH3} have a significantly smaller range of possible abundances. This makes it much harder to be able to link the detection of nitrogen species to the planet's N/H ratio and where it may have formed, especially due to the current scarcity in nitrogen species detected and \ce{HCN} being below the detectable limit regardless of planetary temperature.

\section{A comparison with retrieved abundances in hot Jupiters' atmospheres}
\label{secF:Section5}

In this section we examine an ensemble of molecular abundances measured in the atmospheres of hot Jupiters and compare them to our model predictions. We also perform a case study on the hot Jupiter HD 209458b to investigate whether our models can constrain where the planet formed and how it migrated. 

\begin{table}
\renewcommand{\arraystretch}{1.5}
    \begin{tabular}{c|c|c|c}
    \hline
        
        Planet Name   & $\mathrm{M_P}$ ($\mathrm{M_J}$) & $\mathrm{T_{eq}}$ (K) & $\mathrm{log(X_{\ce{H2O}})}$\\ \hline
        WASP-107b &0.12&740& $-2.87^{+0.95}_{-0.73}$\\
        HD 189733b &1.14&1200&$ -4.66^{+0.35}_{-0.33}$\\
        KELT-11b &0.20&1300& $-3.60^{+0.60}_{-0.70}$\\
        HAT-P-1b &0.53&1320& $-2.54^{+0.75}_{-0.67}$\\
        WASP-43b &2.03&1440& $-3.68^{+0.92}_{-0.88}$\\
        HD 209458b &0.69&1450&$ -4.54^{+0.33}_{-0.27}$\\
         WASP-17b &0.51&1740& $-3.84^{+1.27}_{-0.51}$\\
        WASP-19b &1.14&2050& $-3.43^{+0.47}_{-0.52}$\\
        
    \hline
    \end{tabular}
    \caption[The detected water abundances of eight hot Jupiters]{A table of the \ce{H2O} abundances detected in hot Jupiters with equilibrium temperatures between $1000\,\mathrm{K}$ and $2000\,\mathrm{K}$. The values for KELT-11b come from \protect\cite{Changeat2020}, while all the others were drawn from \protect\cite{Welbanks2019}.}
    \label{tab:Waterabun}
\end{table}

\subsection{\texorpdfstring{\ce{H2O}}{H2O} abundances}

\ce{H2O} is one of the most measured molecule in hot Jupiter atmospheres \citep[e.g.][]{Madhu2014,Kreidberg2014,Barstow2017,Pinhas2019,Welbanks2019}. 
In Table \ref{tab:Waterabun} we show 8 hot Jupiters whose \ce{H2O} abundances have been determined to within an order of magnitude. The values for these planets were mainly drawn from \cite{Welbanks2019}, but similar and generally consistent values for these planets can be found in other works too \citep[e.g.][]{Tsiaras2018, Pinhas2019,Min2020}. These mass, radius and stellar insolation of these planets vary greatly, compared to the single values we have chosen for these parameters in this work. However, these parameters primarily change the way that diffusion or photochemistry affects the abundance of species within the planets atmosphere. As seen in Figures \ref{fig:Kzzcheck} and \ref{fig:UVcheck}, the strength of the $K_{zz}$ profile and incident UV irradiation have no significant affect on the abundance of water, and so we can justify using our model to examine only the water abundance of these planets. All eight planets have their \ce{H2O} abundance fall within the ranges predicted by our models. However, HD 189773b and HD 209458b have notably lower \ce{H2O} abundances than any of the other planets. While the other planets could have formed and migrated by any of the mechanisms discussed in the previous section, our models suggest that HD 189733b and HD 209458b would have had to have formed further out to obtain their current water abundances. Which mechanism is still not precisely determined, but either core accretion, with or without chemical enrichment, beyond the \ce{CO2} snowline, or gravitational instability beyond the \ce{CO} snowline are the only ways to produce such low water mixing ratios within our models.

\begin{table*}
\renewcommand{\arraystretch}{1.5}
    \begin{tabular}{c c c c c}
    Molecule   & Emission & Transmission & Transmission \\
    & (\citealt{Gandhi2019}) & (\citealt{Welbanks2019}) & (\citealt{Giacobbe2021}) & Matching Formation Models\\\hline

        \ce{H2O}   & ${-4.11}^{+0.91}_{-0.3}$ & ${-4.54}^{+0.33}_{-0.27}$ & $-3.92$ & All\\
        \ce{CO}    & ${-2.16}^{+0.99}_{-0.47}$ &&$-2.85$& All\\
        \ce{CH4} &&&$-2.33$ & $\sim$ CE-B or $\sim$ GI-B   \\
        \ce{HCN}&&& $-4.07$ & $\sim$ CA-DF-B or $\sim$ CE or $\sim$ GI-B  \\
        \ce{NH3}&&& $-3.89$& $\sim$ All\\
        C/O ratio & $0.99^{+0.01}_{-0.02}$ &&& CE or GI-B\\

    \end{tabular}
    \caption[Molecular abundances retrieved for HD 209458b]{A summary of the retrieved values of molecular abundances in the atmosphere of HD 209458b from a number of recent studies. The matching formation models column refers to the formation models we compared to in Section \ref{secF:Section4}.  The abbreviations are: CE: Chemical Enrichment. CE-B: Chemical Enrichment beyond the \ce{CO2} snowline. These models came from \protect\cite{Booth2017}. CA-ID: Core Accretion with In Disk migration. CA-DF-B: Core Accretion with Disk-Free Migration from beyond the \ce{CO2} snowline. GI: Gravitational Instability. GI-B: Gravitational Instability between the \ce{CO2} and \ce{CO} snowlines. These models came from \protect\cite{Madhu2014}.}
    \label{tab:HD209Case}
\end{table*}

\subsection{Case study HD 209458b}
HD 209458b is one of the most studied hot Jupiters, with multiple molecules potentially detected within its atmosphere. Here we use the most recent abundance estimates for this planet. In particular, \ce{H2O}, \ce{CO} and the C/O ratio on the day-side of the planet have been retrieved by \cite{Gandhi2019}, and at the day-night terminator we can find values for the \ce{H2O} abundance in \cite{Welbanks2019} and the \ce{H2O}, \ce{CO}, \ce{CH4}, \ce{HCN} and \ce{NH3} abundance from \cite{Giacobbe2021}. These values are summarised in Table \ref{tab:HD209Case}. Based upon General Circulation Models (GCM) of HD 209458b (\cite{Shownman2009}), we can use our $2000\,\mathrm{K}$ model as an approximation to the day-side of the planet, and our $1000\,\mathrm{K}$ model as an approximation to the terminator of the planet. One limitation of using our $1000\,\mathrm{K}$ model for the terminator is that the $K_{zz}$ profile would likely behave differently at the terminator compared to the sub-stellar point due to the temperature gradient at terminator. However, for our 1-D model this approximation should be sufficient for initial estimates. Comparing the retrieved values from the day-side emission spectra in \cite{Gandhi2019} to our results, we see that the detected values for \ce{H2O} and \ce{CO} fall within the range of any of the formation models we considered. However, their retrieved C/O ratio of approximately 1 was only possible in one of the scenarios we investigated: Chemical enrichment via pebble drift, as discussed in \cite{Booth2017}. 

Next we compare our results to the retrieved abundances from the terminator transmission spectra of HD 209458b in \cite{Giacobbe2021}. Both the \ce{H2O} and \ce{CO} results are on the boundary of being possible with any formation model, but they do favour core accretion and chemical enrichment beyond the \ce{CO2} snowline or any formation location by gravitational instability. All of the other abundances of retrieved molecules from \cite{Giacobbe2021} are higher than any our models predict. However, this difference is generally no more than half an order of magnitude, and so we will compare to the formation pathways that produce results closest to the retrieved values. It is likely that these differences arise from either margins of error within the retrieved values, or simplifications within our own models when trying to model the complex chemistry of hot Jupiters. The retrieved value for methane in \cite{Giacobbe2021} is approximately half an order of magnitude above our values of \ce{CH4} for chemical enrichment beyond the \ce{CO2} snowline and gravitational instability between the \ce{CO2} and \ce{CO} snowlines. However, it is several orders of higher than any other formation pathway, thus suggesting that for \ce{CH4}, HD 209458b was likely to form via chemical enrichment beyond the \ce{CO2} snowline or gravitational instability between the \ce{CO2} and \ce{CO} snowlines. The retrieved \ce{HCN} abundance lies above any of our models predictions, but is within half an order of magnitude of values predicted by core accretion and disk-free migration beyond the \ce{CO2} snowline, any location of chemical enrichment, or gravitational instability between the \ce{CO2} and \ce{CO} snowlines. Lastly, the \ce{NH3} results from \cite{Giacobbe2021} are also higher than any of our modelled results, however this could be accounted for by variations in the N/H ratio. Regardless, any formation model could be responsible for producing the observed \ce{NH3} values. 

Overall, the likely formation location and migration pathway for HD 209458b is most consistent with forming between the \ce{CO2} and \ce{CO} snowlines by gravitational instability, since this has the closest fit with every detected feature. However, it is also possible the that HD 209458b formed beyond the \ce{CO2} snowline and was then chemically enriched by pebble accretion. 
\section{Summary and Discussion} 

\label{secF:Conclusion} 

In this work we have explored the carbon, oxygen and nitrogen compositional parameter space for hot Jupiters, and seen how these predicted abundances may relate to their formation location. By expanding upon our previous work in \cite{Hobbs2019} we have run a suite of 1500 models of our chemical kinetics code. We did this for a generic hot Jupiter orbiting at two possible distances from a sun-like star, encompassing a range of atmospheric elemental compositions between 0.1x and 10x the solar values for carbon, oxygen and nitrogen. We compare our parameter space to the formation and migration models of three previous works (\citealt{Madhu2014, Booth2017, Turrini2021}) to create a framework of how the abundance of six of the major species in a hot Jupiter atmosphere might be linked to where the planet formed.

We find that, as expected, there is a large degree of degeneracy when trying to link a planet's atmospheric composition to its formation location, however we do obtain some insights that may assist in narrowing down the history of a hot Jupiter. Using the relationship between formation and metalicity shown in \cite{Madhu2014} we find that low \ce{CO2} and \ce{H2O} abundances can be expected only in planets forming beyond the \ce{CO2} snowline. We also find that low \ce{CH4} abundances on the cooler hot Jupiters are only expected for a planet forming within the \ce{CO2} snowline or by gravitational instability beyond the \ce{CO2} snowline. Additionally, we find that our models predict high \ce{HCN} abundance for planets that formed beyond the \ce{CO2} snowline via core accretion or by gravitational instability between the \ce{CO2} and \ce{CO} snowlines, before undergoing disk-free migration. 


We also considered the possibility of chemical enrichment of hot Jupiters as presented in \cite{Booth2017}. We find the best way to determine if a hot Jupiter had been chemically enriched was via an elevated \ce{CH4} abundance within the enriched planet's atmosphere. The formation location of a chemically enriched planet could best be identified via the \ce{CO} or \ce{CO2} abundance, both of which would be lower in a planet that formed beyond the \ce{CO2} snowline.

Using the work of \cite{Turrini2021}, we investigated whether having the N/H ratio as a chemical parameter could allow us to gain insights into where a hot Jupiter formed. In general we find that this is not the case. Species that do not contain nitrogen are generally unaffected by changes to the N/H ratio. Additionally, the range of N/H ratios predicted by \cite{Turrini2021} only covers between 1 < [N/H] < 3, which does not have a very significant impact even on those species that do contain nitrogen.

We compared an ensemble of retrieved \ce{H2O} abundances on hot Jupiters from \cite{Welbanks2019} to the \ce{H2O} abundances we predicted using our models. We found that all the detected abundances lay within our predictions but, with only the \ce{H2O} abundance available, in most cases it was not possible to distinguish which specific formation mechanism may have lead to that abundance. However, in two cases (HD 189733b and HD 209458b) where the \ce{H2O} abundances are significantly sub-solar \citep{Madhu2014a,Barstow2017, Pinhas2019, Welbanks2019}, our models suggest their formation beyond the \ce{CO2} or \ce{CO} snowlines followed by disk-free migration. Our results are consistent with previous studies which used such low \ce{H2O} abundances in hot jupiters to suggest their formation beyond the snowlines and disk free migration \citep[e.g.][]{Madhu2014,Brewer2017,Welbanks2019}. 

Finally we performed a case study on the hot Jupiter HD 209458b, using multiple molecular detections from \cite{Gandhi2019}, \cite{Welbanks2019} and \cite{Giacobbe2021}. Our model predicted formation between the \ce{CO2} and \ce{CO} snowlines followed by disk free migration as the most likely origin for this planet. This inference assumed a formation model based on gravitational instability, from \cite{Madhu2014}, but formation by core accretion at the same location may also be possible. Similarly, core accretion beyond the \ce{CO2} snowline followed by chemical enrichment by pebble drift \citep{Booth2017} also matched with all but one molecular detection, and thus could also be a contender for how HD 209458b was formed. 

There are naturally some caveats to our results based upon the assumptions made within our modelling. The first of these is the common assumption that the metalicity of the protoplanetary disk is solar. Combining our model results with a planet that formed in an environment that had a significantly super or sub solar metalicity would not allow us to accurately trace a planet back to its formation mechanism and location. Additionally, our model used standardised stellar and planetary properties (i.e., The star was solar equivalent, and the planet modelled was always tidally locked and of a constant mass and radius). These values would be needed to run accurate models of a specific planet. Lastly, chemistry due to other elemental species, such as sulfur, has shown to be able to influence the abundances of the species that we have modelled in this work. Thus, the accuracy of these models could be improved by including these additional species.

To improve our ability to predict formation mechanisms further we need additional, more accurate, molecular abundance measurements on hot Jupiters, and to expand the functionality of our code to consider a wider range of planetary and stellar properties. The upcoming launch of satellites such as the James Web Space Telescope (JWST) should help in providing a more accurate picture of the composition of a hot Jupiter. To improve chemical models of hot Jupiters, inclusion of other elements such as sulfur may assist in making the models more accurate representations of their atmospheric chemistry and breaking degeneracies of formation location.

\section*{Acknowledgements}

R.H. and O.S. acknowledge support from the UK Science and Technology Facilities Council (STFC).

\section*{Data availability}
The data underlying this article will be shared on reasonable request to the corresponding author.




\bibliographystyle{mnras}
\bibliography{bibliography}






\bsp	
\label{lastpage}
\end{document}